\documentclass[lettersize, journal]{IEEEtran}

\usepackage{amsmath,amssymb,amsfonts,graphicx,mathtools,nccmath,amsthm,float}
\usepackage{multicol,tikz,cite,array,booktabs,url}
\usepackage{cite}
\usepackage{textcomp}
\usepackage{xcolor}
\usepackage{algorithm}
\usepackage{algpseudocode,algorithmicx}
\usepackage{acronym}
\usepackage[bookmarks,colorlinks]{hyperref}
\usepackage{xfrac}

\graphicspath{{figures/}{./}}
\DeclareGraphicsExtensions{.pdf}

\newcommand{\vect}[1]{\boldsymbol{\mathbf{#1}}}

\newtheorem{Thm}{Theorem}
\newtheorem{Cor}{Corollary}
\newtheorem{Lem}{Lemma}
\newtheorem{Prop}{Proposition}

\DeclareMathOperator{\diag}{diag}

\DeclareMathOperator{\trace}{Tr}
\DeclarePairedDelimiter{\norm}{\lVert}{\rVert}

\algnewcommand{\LineComment}[1]{\State \(\#\) #1}
\algnewcommand\algorithmicinput{\textbf{Set}}
\algnewcommand\Set{\item[\algorithmicinput]}
\algnewcommand\algorithmicinitial{\textbf{Initialize}}
\algnewcommand\Initialize{\item[\algorithmicinitial]}

\let\oldReturn\Return
\renewcommand{\Return}{\State\oldReturn}

\begin{document}

\title{Multi-RIS-Empowered Multiple Access:\\ A Distributed Sum-Rate Maximization Approach}
\author{Konstantinos~D.~Katsanos,~\IEEEmembership{Graduate Student~Member,~IEEE,} Paolo~Di~Lorenzo,~\IEEEmembership{Senior Member,~IEEE,}\\ and
	George~C.~Alexandropoulos,~\IEEEmembership{Senior Member,~IEEE}
\thanks{This work has been supported by the Smart Networks and Services Joint Undertaking (SNS JU) projects TERRAMETA and 6G-DISAC under the European Union’s Horizon Europe research and innovation programme under Grant Agreement No 101097101 and No 101139130, respectively. TERRAMETA also includes top-up funding by UK Research and Innovation (UKRI) under the UK government’s Horizon Europe funding guarantee.}
	\thanks{K. D. Katsanos and G. C. Alexandropoulos are with the Department of Informatics and Telecommunications, National and Kapodistrian University of Athens, 15784 Athens, Greece (e-mails: \{kkatsan, alexandg\}@di.uoa.gr).}
	\thanks{P. Di Lorenzo is with the Department of Information Engineering, Electronics, and Telecommunications, Sapienza University, Italy and the National Inter-University Consortium for Telecommunications (CNIT), Italy (e-mail: paolo.dilorenzo@uniroma1.it).}
}

\maketitle

\begin{abstract}
The plethora of wirelessly connected devices, whose deployment density is expected to largely increase in the upcoming sixth Generation (6G) of wireless networks, will naturally necessitate substantial advances in multiple access schemes. Reconfigurable Intelligent Surfaces (RISs) constitute a candidate 6G technology capable to offer dynamic over-the-air signal propagation
programmability, which can be optimized for efficient non-orthogonal access of a multitude of devices. In this paper, we study the downlink of a wideband communication system comprising multiple multi-antenna Base Stations (BSs), each wishing to serve an associated single-antenna user via the assistance of a Beyond Diagonal (BD) and frequency-selective RIS. Under the assumption that each BS performs Orthogonal Frequency Division Multiplexing (OFDM) transmissions and exclusively controls a distinct RIS, we focus on the sum-rate maximization problem and present a distributed joint design of the linear precoders at the BSs as well as the tunable capacitances and the switch selection matrices at the multiple BD RISs. The formulated non-convex design optimization problem is solved via successive concave approximation necessitating minimal cooperation among the BSs. Our extensive simulation results showcase the performance superiority of the proposed cooperative scheme over non-cooperation benchmarks, indicating the performance gains with BD RISs via the presented optimized frequency selective operation for various scenarios.
\end{abstract}


\section{Introduction} \label{Sec:Intro}
The upcoming sixth Generation (6G) of wireless networks is envisioned to substantially increase data throughput, improve energy and spectral efficiencies, enhance reliability, decrease latency, and boost massive connectivity, but also inaugurate wireless artificial intelligence \cite{Masaracchia_DT_2023_all,letaief2021edge_all,6Gdisac} and multi-functionality \cite{CMY+24_all,FDmimo6G}. New types of wirelessly connected devices, end users, and the internet of things~\cite{IoT} will further complicate multiple access necessitating simultaneous service delivery within the same communication resources (e.g., time, frequency, power, and space). Reconfigurable Intelligent Surfaces (RISs)~\cite{risTUTORIAL2020,pan2022overview_all,pan2021reconfigurable_all,RIS_Overview_all,RISoverview2023_all} are expected to constitute one of those new types of devices having lately attracted considerable interest from both academia and industry for over-the-air signal propagation programmability~\cite{WavePropTCCN_all}, targeting a multitude of metrics and applications (e.g., coverage extension even in non-line-of-sight environments~\cite{huang2019reconfigurable_all}, energy efficiency~\cite{RIS_amp}, and computation~\cite{RIS_comp}, as well as localization~\cite{RIS_loc}, sensing~\cite{alexandropoulos2023hybrid}, and their integration~\cite{RIS_ISAC}) in a low-cost and low-power manner. 

The potential of RISs to revolutionize wireless connectivity has recently gave birth to the smart wireless environments paradigm\cite{RIS_challenges_all,Alexandropoulos2022Pervasive}, according to which multiple adequately orchestrated RISs are tasked to realize performance-boosted areas~\cite{RIS_rise6g,AOC+24_all}. To this end, there have appeared multiple studies focusing on efficient designs for multi-RIS-empowered wireless systems, targeting, for example, coverage-extending beamforming in high frequencies~\cite{multiRIS_drl}, localization~\cite{AVW22a} and mapping~\cite{HFW+23}, as well as non-orthogonal multiple access \cite{RIS_ngma,ZLZ+23}. The RIS technology has been also lately adopted for Orthogonal Frequency Division Multiplexing (OFDM) systems operating under frequency-selective channels, resulting in significant improvements for various performance objectives~\cite{LZA+20_all,YZZ+20_all}. In \cite{ZD21}, multi-carrier transmissions aided by multiple distributed RISs in a wideband cell-free wireless network with multiple receivers were optimized to maximize the weighted achievable sum rate. The authors in \cite{HXS+21_all} presented a cooperative multi-RIS-assisted transmission scheme for millimetter-wave multi-antenna OFDM systems, which was combined with a delay matching method to reduce the overhead needed for the estimation of multipath channels. Adopting the energy efficiency as the design metric for a wideband RIS-aided cell-free network and using a realistic energy-consumption model in \cite{SMH+22_all}, a joint active and passive beamforming scheme was presented. However, the latter studies ignored the frequency selective behavior of realistic RIS unit elements~\cite{RIS_Scattering_all} and treated the RIS as it would operate in narrowband communications. This comes to contrast with  recent research on wideband RISs~\cite{katsanos2022wideband_all,li2021_wideband_practical_all}, acording to which the frequency response of an RIS metamaterial element is dictated by a Lorentzian-type form, which if ignored leads to notably misleading performance evaluations. Taking into account the frequency selectivity of an RIS, the importance of index modulation in OFDM systems was highlighted in \cite{HBC23_all}, where it was shown that the bit error rate performance can be improved. In \cite{HRC+23_all}, the authors introduced the concept of the optimal phasor in an effort to reduce the frequency selectivity of a wideband RIS. All in all, it becomes apparent that it is of paramount importance to consider accurate wideband models when designing multi-RIS-empowered wireless systems.

A recent advancement on the RIS technology deals with Beyond Diagonal (BD) metasurfaces implementing interconnections between pairs of adjacent phase-tunable metamaterials~\cite{LSN+23_all}, which effectively leads to engineered mutual coupling~\cite{RIS_engineeredMC}. Those connections can be static group/fully-connected or dynamically group-connected resulting naturally in a non-diagonal phase response matrix~\cite{RIS_tacit}. It was shown via simulations in \cite{LSC22} that, in the passive operation mode, a fully- and group-connected RIS can effectively outperform a typical diagonal RIS~\cite{huang2019reconfigurable_all} in terms of the achievable sum-rate performance. To further increase the potential of a BD RIS, while avoiding the imposed complexity of its configuration, a dynamic grouping strategy was proposed in \cite{LSC23}, according to which the RIS unit elements adapt to the available channel state information and are dynamically grouped into several subsets. This approach yielded improved performance when compared to fixed element grouping architectures. Very recently, in \cite{DRM+24_all}, the frequency-dependent behavior of BD RISs was investigated in the framework of multi-band multi-cell Multiple-Input Multiple-Output (MIMO) systems, showcasing that BD RISs can outperform their diagonal counterparts. Nevertheless, to the best of the authors' knowledge, it has not yet been investigated how multiple BD and frequency-selective RISs perform in multiple access setups with OFDM.

The available literature on multi-RIS-empowered smart wireless environments indicates that programmable metasurfaces are capable to offer increased performance especially in cases where multiple users request simultaneously high levels of quality of service. However, the performance of the interference channel with multiple BD RISs exhibiting frequency selective response has not being yet reported, a fact that motivated this work. In particular, in this paper, we consider multiple distributed frequency-selective BD RISs, each controlled by a distinct multi-antenna Base Station (BS) performing downlink OFDM transmissions to its assigned single-antenna user. Differently from our preliminary work~\cite{KDA22a}, we focus on BD RIS structures and present a more efficient optimization approach for the involved system parameters, which is shown to provide increased performance. The main contributions of this paper are summarized as follows.
\begin{itemize}
    \item We consider multiple distributed BD RISs and present a tractable analytical form for their common frequency response, which is parameterized over the tunable capacitances of the equivalent circuits for each of their metamaterial unit elements. We assume that each BD RIS structure comprises an array of binary states/switches, being deployed to interconnect all of its unit elements providing extra degrees of freedom for optimization.    
    \item Focusing on the sum-rate maximization as the objective of the considered multi-RIS-empowered multiple access system, we formulate a novel challenging mixed-integer non-convex optimization problem, aiming to jointly design the linear precoding vector at each multi-antenna BS as well as the reflective beamforming vector and configuration of the switches at each BD RIS structure.
    \item By decomposing the considered optimization problem into local sub-problems that can be solved in parallel, we present a provably convergent 
    distributed algorithm running on each distinct BS and requiring minimal cooperation via pricing matrices exchange among them.      
    \item We present extensive simulation results for the proposed distributed multi-RIS-empowered multi-access system design, and its performance is compared with benchmark schemes. It is showcased that our minimal BS cooperation improves significantly the achievable sum-rate performance. It is also demonstrated that, with the proposed design, both the phase shifts and the amplitudes of each BD RIS's reflection coefficients are optimized at each frequency bin across the entire operation bandwidth. 
\end{itemize}

The remainder of the paper is organized as follows. Section~\ref{Sec:Sys_Model_and_Problem} includes our system model, the models for the BD RIS response and the received signal, as well as our design problem formulation. Section~\ref{Sec:Design_SCA} presents our distributed scheme for sum-rate maximization, while Section~\ref{Sec:Numerical} discusses our extensive results for the system performance. Finally, Section \ref{Sec:Conclusion} accommodates the paper's concluding remarks. 

\textit{Notations:} Vectors and matrices are denoted by boldface lowercase and boldface capital letters, respectively. The transpose, conjugate, Hermitian transpose, and inverse of $\mathbf{A}$ are denoted by $\mathbf{A}^{\rm T}$, $\mathbf{A}^*$, $\mathbf{A}^{\rm H}$, and $\mathbf{A}^{-1}$ respectively, while $\mathbf{I}_{n}$ ($n\geq2$) and $\mathbf{0}_{m \times n}$ ($m\geq2$) are the $n\times n$ identity matrix and the $m\times n$ zeros' matrix, respectively. ${\rm Tr}(\mathbf{A})$, $\norm{\mathbf{A}}_{\rm F}$, and $\norm{\mathbf{a}}$ represent $\mathbf{A}$'s trace and Frobenius norm, and $\mathbf{a}'s$ Euclidean norm, respectively, while notation $\mathbf{A}\succ\vect{0}$ ($\mathbf{A}\succeq\vect{0}$) means that the square matrix $\mathbf{A}$ is Hermitian positive definite (semi-definite). $[\mathbf{A}]_{i,j}$ is the $(i,j)$-th element of $\mathbf{A}$, $[\mathbf{a}]_i$ is $\mathbf{a}$'s $i$-th element, and ${\rm diag}\{\mathbf{a}\}$ denotes a square diagonal matrix with $\mathbf{a}$'s elements in its main diagonal. $\operatorname{vec}(\vect{A})$ indicates the vector which is comprised by stacking the columns of $\vect{A}$, and $\operatorname{vec}_{\rm d}(\vect{A})$ denotes the vector obtained by the diagonal elements of the square matrix $\vect{A}$. $\nabla_{\mathbf{a}}f$ denotes the Euclidean gradient vector of a scalar function $f(\cdot)$ along the direction indicated by $\mathbf{a}$. $\mathbb{C}$ represents the complex number set, $\jmath \triangleq \sqrt{-1}$ is the imaginary unit, $|a|$ denotes the amplitude of the complex scalar $a$, and $\Re(a)$ returns $a$'s real part. $<\!\mathbf{A}_1,\mathbf{A}_2\!>\,\triangleq \Re\{\trace(\mathbf{A}_1^{\rm H} \mathbf{A}_2)\}$ denotes the inner product between matrices $\mathbf{A}_1$ and $\mathbf{A}_2$ of suitable dimensions. $\mathbb{E}[\cdot]$ is the expectation operator and $\mathbf{x}\sim\mathcal{CN}(\mathbf{a},\mathbf{A})$ indicates a complex Gaussian random vector with mean $\mathbf{a}$ and covariance matrix $\mathbf{A}$. Finally, $\mathcal{O}(f(x))$ represents the Big-O notation for the function $f(x)$.


\section{System Model and Problem Formulation} \label{Sec:Sys_Model_and_Problem}
In this section, we present our multi-RIS-empowered interference channel together with the considered wideband model for the element response of each BD RIS and the resulting received signal model at each user. The formulation of our joint design objective for the precoders at the multiple BSs as well as the tunable capacitances and the switch selection matrices at the BD RISs is also introduced.

\subsection{System Model} \label{Sec:Sys_Model}
\begin{figure}[!t]
	\centering
	\includegraphics[width=3.45in]{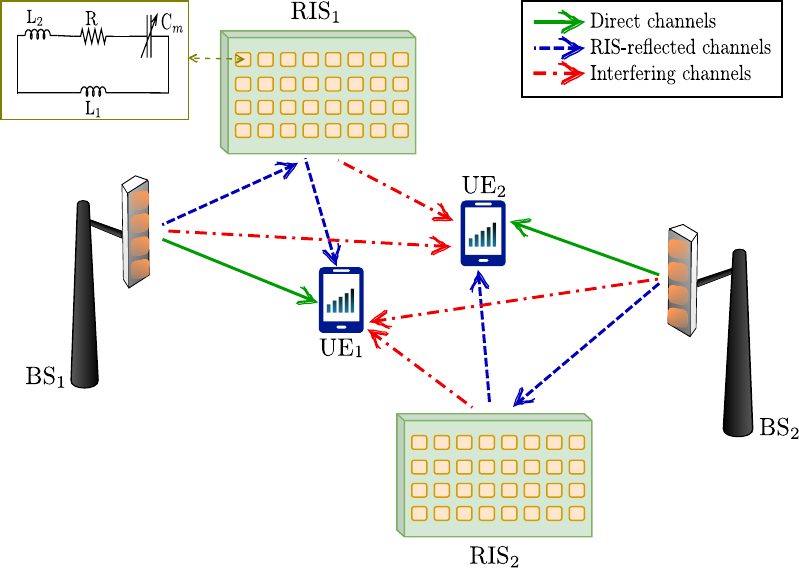}
	\caption{\small{The considered multi-RIS-empowered multi-access wireless system comprising $Q$ multi-antenna BSs each controlling a separate BD RIS and communicating in the downlink direction with a separate single-antenna UE; the case of $Q=2$ is illustrated.}}
	\label{fig:System_Model}
\end{figure}

We consider a multi-RIS-empowered multi-user wireless communication system comprising of $Q$ multi-antenna BSs each wishing to communicate in the downlink direction with a single-antenna User Equipment (UE) via a dedicated solely reflecting RIS~\cite{RISoverview2023_all}, as depicted in Fig.~\ref{fig:System_Model}. We assume that each BS sends information to its exclusively associated UE using OFDM in a common set of physical resources, e.g., time and bandwidth. Thus, there exist $Q$ communicating BS-UE pairs, each modeled as the superposition of a direct BS-UE link and a BS-RIS-UE link realized via the RIS-enabled tunable reflection. Each RIS, comprising $M$ passive reflecting elements connected to a dedicated controller responsible to dynamically adjust their electromagnetic responses for desired signal reflection, is assumed to be controlled by its solely owned BS and to be placed either closely to it or near to its corresponding UE for optimum peformances~\cite{alexandropoulos2022smart_all,Moustakas_Cap2023}.

We consider a BD RIS structure~\cite{li2022_nonDiag_switches_all} according to which an $M\times M$ array of ON/OFF-state switches is deployed to interconnect all RIS unit elements. Specifically, an ON-state at the switch in the position $(i,j)$ (with $i,j=1,2,\ldots,M$) of the switch array indicates that the signal impinging on the $i$-th metamaterial element will be guided to and tunably reflected by the $j$-th element. This behavior can be mathematically expressed by a selection matrix $\vect{S}_q \in \{0,1\}^{M\times M}$ with $q=1,2,\ldots,Q$, whose role is to indicate the switch array selection process at each $q$-th RIS. In particular, each $\vect{S}_q$ is a binary-valued selection matrix (i.e., $[\vect{S}_q]_{i,j} \in \{0,1\}$) which by definition needs to satisfy the property of having only one non-zero value per row and column simultaneously and, thus, constitutes an extra design parameter, as will be discussed in the sequel. It is noted that a typical diagonal RIS, which does not require switches~\cite{RISoverview2023_all}, is obtained by setting $\vect{S}_q=\mathbf{I}_M$.

According to the deployed OFDM scheme, the total bandwidth is equally split into $K$ orthogonal Sub-Carriers (SCs). Let $\vect{w}_{qk} \in \mathbb{C}^{N \times 1}$, with $k=1,2,\ldots,K$ and $N$ denoting the number of antenna elements at each BS, represent the linear precoding vector at each $q$-th BS that models the digital spatial processing of its unit-power signal $x_{qk}$ (i.e., $\mathbb{E}\{|x_{qk}|^2\} = 1$) before transmission. We assume that the total transmit power available at each $q$-th BS is given by $P_q$. Thus, each precoding vector must satisfy the condition $\sum_{k=1}^K \norm{\vect{w}_{qk}}^2\leq P_q$ for each $q$-th BS. We consider a quasi-static block fading channel model for all channels involved and focus on each particular fading block where all channels in the system remain approximately constant.

\subsection{RIS Element Response} \label{Sec:Freq_Response}
Each of the considered almost passive and solely reflecting BD RISs is equipped with $M$ metamaterial elements arranged in a planar structure. We make the quite general assumption that each $m$-th unit element ($m=1,2,\ldots,M$) of each $q$-th RIS can be characterized as an equivalent resonant circuit comprising a resistor $R$, a tunable capacitor $C_{mq}$, and two inductors $L_1$ and $L_2$ \cite{zhu2013active_impedance_all}. Then, the response of each $m$-th unit element of each $q$-th RIS is given by the ratio between the power of the incident signal and that of the reflected one. This ratio indicates the reflection coefficient which is expressed in the frequency domain as follows:
\begin{equation} \label{eqn:reflect_coeff}
	\phi_{mq}(f,C_{mq}) = \frac{\mathcal{Z}(f,C_{mq}) - \mathcal{Z}_0}{\mathcal{Z}(f,C_{mq}) + \mathcal{Z}_0},
\end{equation}
where $\mathcal{Z}(f,C_{mq})$ denotes the characteristic impedance of the equivalent circuit which is given for $\kappa\triangleq2\pi$ by
\begin{equation} \label{eqn:characteristic_impedance}
	\mathcal{Z}(f,C_{mq}) = \frac{\jmath \kappa f L_1\left(\jmath \kappa f L_2 + R + \frac{1}{\jmath \kappa f C_{mq}}\right)}{\jmath \kappa f \left(L_1+L_2\right) + R + \frac{1}{\jmath \kappa f C_{mq}}},
\end{equation}
while $\mathcal{Z}_0$ is the free space impedance. It is noted that this reflection coefficient expression is a complicated one, which is hard to deal with, since it is a highly non-linear function of the tunable capacitor $C_{mq}$. In addition, to the best of the authors' knowledge, all relevant studies construct fitting functions to approximate the behavior of the above response, targeting on the simplification of its analysis. However, in this work, we observe that \eqref{eqn:reflect_coeff} can be equivalently transformed into a more tractable form, according to the following proposition.
\begin{Prop} \label{prop:RIS_Freq_Response}
	The frequency response of each $m$-th unit element of each $q$-th RIS can be reformulated as follows:
	\begin{equation} \label{eqn:RIS_freq_response}
		\phi_{mq}(f,C_{mq}) = 1 - \frac{2}{1 + \frac{\mathcal{D}_{mq}(f,C_{mq})}{\mathcal{N}_{mq}(f,C_{mq})}},
	\end{equation}
	where $\mathcal{N}_{mq}(f,C_{mq})$ and $\mathcal{D}_{mq}(f,C_{mq})$ are defined as follows:
	\begin{align}
		\mathcal{N}_{mq}(f,C_{mq}) \!&\triangleq\! 1 \!-\! (\kappa f)^2(L_1 \!+\! L_2)C_{mq} \!+\! \jmath\kappa f R C_{mq}, \label{eqn:numerator_s} \\ 
		\mathcal{D}_{mq}(f,C_{mq}) \!&\triangleq\! \jmath \kappa f\frac{L_1}{\mathcal{Z}_0}\left(1\!-\!(\kappa f)^2 L_2 C_{mq}\!+\!\jmath \kappa f R C_{mq} \right)\!. \label{eqn:denominator_s}
	\end{align}
\end{Prop}
\begin{IEEEproof}
	See Appendix \ref{appx:RIS_Freq_Response}.
\end{IEEEproof}

\subsection{Received Signal Model} \label{Sec:Received_Model}
According to the considered system model, for each $q$-th BS-UE pair, there will be an additional RIS-enabled wireless link, namely, the $q$-th BS-RIS-UE link, through which the signals transmitted by the $q$-th BS are reflected by its owned $q$-th RIS before arriving at the intended $q$-th UE. Let $\vect{H}_{qq,k} \in \mathbb{C}^{M\times N}$ and $\vect{g}_{qq,k} \in \mathbb{C}^{M \times 1}$ denote each $q$-th BS-RIS and each $q$-th RIS-UE channel, respectively, at each $k$-th SC. Each $m$-th phase-tunable unit element of each $q$-th RIS scatters its impinging signals via an independent reflection coefficient $\phi_{mq}$ whose frequency response is given by \eqref{eqn:RIS_freq_response}, where $C_{mq}$ indicates its tunable capacitance value. We define the vector $\vect{\phi}_{q,k} \triangleq [\phi_{1q}(f_k,C_{1q}),\ldots,\phi_{Mq}(f_k,C_{Mq})]^{\rm T} \in \mathbb{C}^{M \times 1}$ as the one including the reflection coefficients of each $q$-th RIS and then formulate the matrices $\vect{\Phi}_{q,k} \triangleq \operatorname{diag}\{\vect{\phi}_{q,k}\}$ $\forall$$q,k$. The baseband received signal at each $q$-th UE at each $k$-th SC (i.e., in the frequency domain) can be mathematically expressed as:
\begin{equation} \label{eqn:received_signal}
	y_{qk} = \vect{f}_{qq,k}^{\rm H} \vect{w}_{qk}x_{qk} + \sum_{j\neq q}^Q \vect{f}_{jq,k}^{\rm H} \vect{w}_{jk} x_{jk} + n_{qk},
\end{equation}
where we have used the following definitions for the end-to-end channels:
\begin{align} 
	\vect{f}_{qq,k}^{\rm H} &\triangleq \vect{h}_{qq,k}^{\rm H} + \vect{g}_{qq,k}^{\rm H}\vect{S}_q\vect{\Phi}_{q,k}\vect{H}_{qq,k}, \label{eqn:total_channels_1}\\
	\vect{f}_{jq,k}^{\rm H} &\triangleq \vect{h}_{jq,k}^{\rm H} + \vect{g}_{jq,k}^{\rm H}\vect{S}_j\vect{\Phi}_{j,k}\vect{H}_{jj,k}.
 \label{eqn:total_channels_2}
\end{align}
In the latter expressions, each $\vect{h}_{jq,k}\in \mathbb{C}^{N \times 1}$ with $q,j=1,2,\ldots,Q$ indicates the direct channel between the $q$-th UE and the $j$-th BS at each $k$-th SC, and $n_{qk} \sim \mathcal{CN}(0,\sigma_q^2)$ represents the Additive White Gaussian Noise (AWGN), which models the thermal noises at the UE receivers.

It can be seen from the received signal model in \eqref{eqn:received_signal} and from Fig.~\ref{fig:System_Model} that the considered multi-RIS-empowered multiple access scheme generates interference signals at each receiving UE. These signals are included in the middle term of \eqref{eqn:received_signal} and refer to the interference links at each UE resulting from any unintended BS and via the reflections from any unintended RIS (e.g., the only intended BS and RIS for the $q$-th UE are the $q$-th BS and the $q$-th RIS). We have assumed that interference coming from any unintended BS to a UE via an intended RIS is negligible due to appropriate RIS placement~\cite{RIS_challenges_all}.   

\subsection{Design Objective} \label{Sec:Prob_Form}
We define the vectors: \textit{i}) $\widetilde{\vect{w}} \triangleq [\vect{w}_1^{\rm T},\dots,\vect{w}_Q^{\rm T}]^{\rm T}$ with $\vect{w}_q \triangleq [\vect{w}_{q1}^{\rm T},\dots,\vect{w}_{qK}^{\rm T}]^{\rm T}$; and \textit{ii}) $\widetilde{\vect{c}}\triangleq [\vect{c}_1^{\rm T},\ldots,\vect{c}_Q^{\rm T}]^{\rm T}$ with $\vect{c}_q \triangleq [C_{1q},\ldots,C_{Mq}]^{\rm T} \in \mathbb{R}^{M \times 1}$; as well as \textit{iii}) the set of matrices $\widetilde{\vect{S}} \triangleq \{\vect{S}_q\}_{q=1}^Q$ including, respectively, the precoding vectors at the $Q$ multi-antenna BSs, the tunable capacitances, and the switch selection matrices at the $Q$ BD RISs. By treating the Multi-User Interference (MUI) term in \eqref{eqn:received_signal} as an additional source of noise (colored noise), the achievable sum-rate performance in bits per second per Hz (bits/s/Hz) for each $q$-th UE can be expressed as the following function of the triplet with the tunable parameters $(\widetilde{\vect{w}},\widetilde{\vect{c}},\widetilde{\vect{S}})$: 
\begin{equation} \label{eqn:sum_rate_q}
\mathcal{R}_q\left(\widetilde{\vect{w}},\widetilde{\vect{c}},\widetilde{\vect{S}}\right) = \sum_{k=1}^K \log_2\left( 1 + \frac{|\vect{f}_{qq,k}^{\rm H}\vect{w}_{qk}|^2}{\sigma_q^2 + \sum_{j \neq q}^Q |\vect{f}_{jq,k}^{\rm H}\vect{w}_{jk}|^2} \right),
\end{equation}
where the dependence on $\widetilde{\vect{c}}$ and $\widetilde{\vect{S}}$ is implied via the composite channels $\vect{f}_{qq,k}$ and $\vect{f}_{jq,k}$ in \eqref{eqn:total_channels_1} and \eqref{eqn:total_channels_2}, respectively. It is noted that we have intentionally removed the necessary multiplicative factor $\frac{1}{K}$ that should appear outside the summation in~\eqref{eqn:sum_rate_q}, since it will not affect our optimization formulation and solution that will be described in the sequel.

In this paper, we aim to maximize the achievable sum-rate performance of the proposed multi-RIS-empowered multi-access wireless system in a distributed manner. In particular, we consider the following optimization problem:
\begin{align*}
	\mathcal{OP}: \,\max_{\widetilde{\vect{w}},\widetilde{\vect{c}},\widetilde{\vect{S}}} \, & \quad \sum_{q=1}^Q \mathcal{R}_q\left(\widetilde{\vect{w}},\widetilde{\vect{c}},\widetilde{\vect{S}}\right) \\
	\text{s.t.} & \quad \sum_{k=1}^{K} \norm{\vect{w}_{qk}}^2 \leq P_q, \;\;\vect{S}_q \in \mathcal{S},\;\;\forall q = 1,2,\ldots,Q,\\
	& \quad C_{\min} \leq [\vect{c}_q]_m \leq C_{\max},  \;\; \forall m=1,2,\ldots,M,
\end{align*}
where $\mathcal{S} \triangleq\left\{ \vect{S}\in \{0,1\}^{M\times M}:\vect{S}\vect{1} = \vect{1}, \vect{S}^{\rm T}\vect{1}= \vect{1} \right\}$ indicates the feasible set for the switch selection matrices at the BD RISs, while $C_{\min}$ and $C_{\max}$ represent respectively the minimum and maximum allowable values for the RIS tunable capacitances according to circuital characteristics. To solve $\mathcal{OP}$ (which is provably an NP-hard problem \cite{scutari2013decomposition_all}) in a distributed manner, we assume that each $q$-th BS possesses the channel gain matrices included in $\vect{f}_{qq,k}$ in \eqref{eqn:total_channels_1} $\forall$$k$ (see \cite{RIS_Overview_all} and references therein for wideband channel estimation techniques in RIS-aided communication systems) and deploys a Successive Concave Approximation (SCA) algorithmic framework similar to \cite{scutari2013decomposition_all}, as will be presented in the following section. This framework enables the efficient decomposition of $\mathcal{OP}$ into $Q$ sub-problems that can be iteratively solved in parallel by each individual BS, requiring only minimal message exchanges among their relevant processing units.

\section{Distributed Sum-Rate Maximization} \label{Sec:Design_SCA}
Let $\vect{X}_q \triangleq \{\vect{w}_q,\vect{c}_q,\vect{S}_q\}$ be the set of variables associated with the precoding vector as well as the tunable capacitances and the switch selection matrix at each $q$-th BS-RIS-UE triplet, which we will henceforth refer to as the $q$-th user. Let also $\vect{X}_{-q}$ be the set of all other users' variables except the $q$-th one. The objective function in $\mathcal{OP}$ is non-concave, due to the presence of MUI and the coupling between the optimization variables, which makes the design of a distributed solution procedure even harder. However, we first note that the sum-rate objective in $\mathcal{OP}$ can be re-written in the following form:
\begin{equation} \label{eqn:total_rate}
	\overline{\mathcal{R}}(\vect{X}_q,\vect{X}_{-q})\,\triangleq\,\mathcal{R}_q(\vect{X}_q,\vect{X}_{-q})+\sum_{j\neq q}^Q \mathcal{R}_j(\vect{X}_q,\vect{X}_{-q}).
\end{equation}
The structure in (\ref{eqn:total_rate}) naturally leads to a decomposition scheme having the following form: \textit{i}) at every iteration $t$, the first term $\mathcal{R}_q(\vect{X}_q,\vect{X}_{-q})$ is replaced by a surrogate function, denoted as $\widetilde{\mathcal{R}}_q(\vect{X}_q,\vect{X}^t)$, which depends on the current iterate $\vect{X}^t$; and \textit{ii}) the term $\sum_{j\neq q}^Q \mathcal{R}_j(\vect{X}_q,\vect{X}_{-q})$ is linearized around $\vect{X}_q^t$. The rationale is to locally approximate $\mathcal{OP}$ through a sequence of strongly concave optimization problems, amenable for distributed implementation at each $q$-th BS, while preserving optimality of the final solution \cite{scutari2013decomposition_all}. Thus, the proposed updating scheme for distributedly solving $\mathcal{OP}$ reads as: at every algorithmic iteration $t$, each user $q$ solves the following surrogate optimization problem:
\begin{equation*}\label{eqn:Surrogate_problem}
	\mathcal{OP}_1:\quad\widehat{\vect{X}}_q^t\,=\, \arg\max_{\vect{X}_q \in \mathcal{X}_q} \, \widetilde{\mathcal{R}}_q(\vect{X}_q;\vect{X}^t)+<\boldsymbol{\Pi}^t_q,\vect{X}_q -\vect{X}_q^t>, 
\end{equation*}
where $\mathcal{X}_q$ denotes the feasible set combining all constraints of $\mathcal{OP}$, while the local surrogate function $\widetilde{\mathcal{R}}_q(\cdot;\cdot)$ has the following structure\footnote{We henceforth assume that $\vect{f}_{qq,k}$ encapsulates the optimization variables $\vect{c}_q^t$ and $\vect{S}_q^t$, but we omit the superscript $t$ for clarity of presentation.}:
\begin{align}
	\begin{split}		
		\widetilde{\mathcal{R}}_q(\vect{X}_q;\vect{X}^t) &\triangleq \sum_{k=1}^K \log_2 \left( 1 + \frac{|\vect{f}_{qq,k}^{\rm H}\vect{w}_{qk}|^2}{\sigma_q^2 + \sum_{j\neq q}^Q|\vect{f}_{jq,k}^{\rm H}\vect{w}_{jk}^t|^2} \right) \\
		&\phantom{ = {}} -\frac{\tau}{2}\Big(\norm{\vect{w}_q - \vect{w}_q^t }^2 +\norm{\vect{c}_q - \vect{c}_q^t}^2 \\
		&\phantom{ = {}}+ \norm{ \vect{S}_q - \vect{S}_q^t}_{\rm F}^2\Big) + <\vect{\gamma}_{\vect{c}_q}^t,\vect{c}_q - \vect{c}_q^t>  \\
		&\phantom{ = {}}+ <\vect{\Gamma}_{\vect{S}_q}^t,\vect{S}_q - \vect{S}_q^t>, \label{eqn:first_surrogate}
	\end{split}
\end{align}
with $\tau > 0$ being an appropriately chosen parameter, $\vect{\gamma}_{\vect{c}_q}^t \triangleq \nabla_{\vect{c}_q}\mathcal{R}_q(\vect{X}_q,\vect{X}_{-q}^t)\vert_{\vect{c}_q=\vect{c}_q^t}$ and, similarly, $\vect{\Gamma}_{\vect{S}_q}^t \triangleq \nabla_{\vect{S}_q}\mathcal{R}_q(\vect{X}_q;\vect{X}^t)\vert_{\vect{S}_q=\vect{S}_q^t}$. In $\mathcal{OP}_1$, $\vect{\Pi}_q^t$ is defined as follows:
\begin{equation}\label{eqn:pricing_def}
    	\vect{\Pi}_q^t \triangleq \sum_{j\neq q}^Q \nabla_{\vect{X}_q} \mathcal{R}_j(\vect{X}_q,\vect{X}_{-q})\Big\vert_{\vect{X}_q=\vect{X}_q^t}, 
\end{equation}
where the single term $\nabla_{\vect{X}_q} \mathcal{R}_j(\vect{X}_q,\vect{X}_{-q})\Big\vert_{\vect{X}_q=\vect{X}_q^t}$ is frequently referred to in the literature as the \textit{pricing vector}~\cite{scutari2013decomposition_all}, i.e., a term that quantifies how the allocation of resources to user $q$ affects the achievable rate of user $j$. Then, our methodology proceeds by solving $\mathcal{OP}_1$ at each $q$-th BS in a distributed fashion for each set of variables $\vect{X}_q$, as explicitly described in the next subsections.

\subsection{Local Linear Precoding Optimization} \label{Sec:Precoder}
Solving $\mathcal{OP}_1$ with respect to the linear precoder $\vect{w}_q$ for the $q$-th BS leads to the following optimization sub-problem:
\begin{align*}
	\begin{split}
		\mathcal{OP}_2: \,\max_{\vect{w}_q} \, & \,\, \sum_{k=1}^K \breve{\mathcal{R}}_{qk}(\vect{w}_{qk}) - \frac{\tau}{2}\norm{\vect{w}_q - \vect{w}_q^t}^2 \\
		&+ \Re \left\{ (\overline{\vect{\pi}}_q^t)^{\rm H}(\vect{w}_q - \vect{w}_q^t) \right\}\\
		\text{s.t.} & \,\, \sum_{k=1}^{K} \norm{\vect{w}_{qk}}^2 \leq P_q,
	\end{split}
\end{align*}
where $\breve{\mathcal{R}}_{qk}$ stands for the logarithmic term in \eqref{eqn:first_surrogate} and $\overline{\vect{\pi}}_q^t \triangleq \sum_{j\neq q}^Q\nabla_{\vect{w}_q} \mathcal{R}_j(\vect{w}_q,\vect{w}_{-q})\Big\vert_{\vect{w}_q=\vect{w}_q^t}$ is the part of the pricing vector associated with $\vect{w}_q$, according to \eqref{eqn:pricing_def}. In particular, the pricing vector is defined as $\overline{\vect{\pi}}_q^t \triangleq [(\overline{\vect{\pi}}_{q1}^t)^{\rm T},\ldots,(\overline{\vect{\pi}}_{qK}^t)^{\rm T}]^{\rm T} \in \mathbb{C}^{KN \times 1}$ and its $k$-th sub-vector $\overline{\vect{\pi}}_{qk}^t$ associated with the corresponding SC is given by the expression:
\begin{equation} \label{eqn:pricing_w_q}
	\overline{\vect{\pi}}_{qk}^t = -\frac{1}{\ln(2)}\sum_{j \neq q}^Q \frac{\operatorname{snr}_{jk}^t}{(1 + \operatorname{snr}_{jk}^t)\operatorname{MUI}_{jk}^t}\vect{f}_{qj,k} \vect{f}_{qj,k}^{\rm H} \vect{w}_{qk}^t
\end{equation}
$\forall$$q=1,2,\ldots,Q$ and $\forall$$k=1,\ldots,K$, where $\operatorname{snr}_{jk}^t$ and $\operatorname{MUI}_{jk}^t$ are respectively the SINR and the MUI-plus-noise power experienced by user $j$, that are generated by the local precoding vector $\vect{w}_{qk}^t$, which are obtained as:
\begin{align}
	\operatorname{snr}_{jk}^t &\triangleq \frac{\left\lvert \vect{f}_{jj,k}^{\rm H} \vect{w}_{jk}^t \right\rvert^2}{\operatorname{MUI}_{jk}^t}, \\ 
	\operatorname{MUI}_{jk}^t &\triangleq \sigma_j^2 + \sum_{q \neq j}^Q \left\lvert \vect{f}_{qj,k}^{\rm H} \vect{w}_{qk}^t \right\rvert^2.
\end{align}
Regrettably, $\mathcal{OP}_2$  persists as a non-concave problem, primarily owing to the logarithmic function involving the quadratic term with respect to $\vect{w}_{qk}$ in $\breve{\mathcal{R}}_{qk}$. Nonetheless, the remaining components of the objective function, namely, the summation of the squared negative norm augmented by the linear term, exhibit concavity, while the constraint set remains convex. To address the challenge posed by the non-concave terms and maintain a valid surrogate function that retains first-order information, we leverage the following Lemma.

\begin{Lem} \label{lem:logarithmic_surrogate}
	The logarithmic term $\breve{\mathcal{R}}_{qk}(\vect{w}_{qk})$ in \eqref{eqn:first_surrogate} can be lower-bounded by the following surrogate function: 
	\begin{equation} \label{eqn:logarithmic_surrogate}
		\begin{aligned}
			\widehat{\mathcal{R}}_{qk}(\vect{w}_{qk}) = -a_{qk}^t\vect{w}_{qk}^{\rm H}\vect{f}_{qq,k}\vect{f}_{qq,k}^{\rm H}\vect{w}_{qk} + 2\Re\{(\vect{b}_{qk}^t)^{\rm H}\vect{w}_{qk}\},
		\end{aligned}
	\end{equation}
	where $a_{qk}^t$ and $\vect{b}_{qk}^t \in \mathbb{C}^{N \times 1}$ are defined as:
	\begin{align}
		a_{qk}^t &\triangleq \frac{1}{\ln(2)}\frac{\lvert \vect{f}_{qq,k}^{\rm H} \vect{w}_{qk}^t \rvert^2}{(\operatorname{MUI}_{qk}^t + \lvert \vect{f}_{qq,k}^{\rm H} \vect{w}_{qk}^t \rvert^2) \operatorname{MUI}_{qk}^t}, \label{eqn:surrog_a} \\
		\vect{b}_{qk}^t &\triangleq \frac{1}{\ln(2)}\frac{1}{\operatorname{MUI}_{qk}^t} \vect{f}_{qq,k} \vect{f}_{qq,k}^{\rm H} \vect{w}_{qk}^t. \label{eqn:surrog_b}
	\end{align}
\end{Lem}
\begin{IEEEproof}
	The proof is delegated in Appendix \ref{appx:logarithmic_surrogate}.
\end{IEEEproof}
According to Lemma~\ref{lem:logarithmic_surrogate}, it can be easily verified that the derived surrogate function $\breve{\mathcal{R}}_{qk}$ shares the same first-order properties with $\widehat{\mathcal{R}}_{qk}$, that is, the following equality holds:
\begin{equation} \label{eqn:obj_surrog_gradients}
    \nabla_{\vect{w}_{qk}} \widehat{\mathcal{R}}_{qk} \Big\vert_{\vect{w}_{qk} = \vect{w}_{qk}^t} = \nabla_{\vect{w}_{qk}} \breve{\mathcal{R}}_{qk} \Big\vert_{\vect{w}_{qk} = \vect{w}_{qk}^t},
\end{equation}
which is important for the convergence properties of the proposed optimization procedure \cite{scutari2013decomposition_all}.

Next, exploiting Lemma~\ref{lem:logarithmic_surrogate} in $\mathcal{OP}_2$, we obtain the local (strongly) concave optimization problem at user $q$: 
\begin{align*}
	\widehat{\mathcal{OP}}_2: \,\max_{\vect{w}_q} & \sum_{k=1}^K \left(\!-\!a_{qk}^t \vect{w}_{qk}^{\rm H}\vect{f}_{qq,k}\vect{f}_{qq,k}^{\rm H}\vect{w}_{qk} + 2\Re\{(\vect{b}_{qk}^t)^{\rm H} \vect{w}_{qk}\}\right)\\
	&- \frac{\tau}{2}\norm{\vect{w}_q - \vect{w}_q^t}^2 + \Re\{ (\overline{\vect{\pi}}_q^t)^{\rm H}(\vect{w}_q - \vect{w}_q^t) \} \\
	\text{s.t.} & \,\, \sum_{k=1}^{K} \norm{\vect{w}_{qk}}^2 \leq P_q.
\end{align*}
To proceed with the solution of $\widehat{\mathcal{OP}}_2$, it suffices to define the block diagonal matrix $\tilde{\vect{F}}_q \triangleq \operatorname{blkdiag}\{a_{qk}^t\vect{f}_{qq,k} \vect{f}_{qq,k}^{\rm H}\}_{k=1}^K$ and the vector $\tilde{\vect{f}}_q \triangleq [(\vect{b}_{q1}^t)^{\rm T},\ldots,(\vect{b}_{qK}^t)^{\rm T}]^{\rm T} \in \mathbb{C}^{KN \times 1}$, allowing to express its objective function, $\mathcal{J}$, after the norm expansion and some algebraic manipulations, as follows:
\begin{equation}
	\mathcal{J}=-\vect{w}_q^{\rm H}\left(\tilde{\vect{F}}_q + \frac{\tau}{2}\vect{I}_{KN} \right)\vect{w}_q + \Re\left\{(\vect{v}_q^t)^{\rm H}\vect{w}_q \right\},
\end{equation}
where $\vect{v}_q^t \triangleq \overline{\vect{\pi}}_q^t + 2\tilde{\vect{f}}_q + \tau \vect{w}_q^t$. Consequently, by noting that $\tilde{\vect{F}}_q \succeq \vect{0}$, it can be easily verified that this compact form of the objective function is concave. Therefore, $\widehat{\mathcal{OP}}_2$ is a (strongly) concave problem, whose Lagrangian function is given by:
\begin{equation}
	\begin{aligned}
		\mathcal{L}(\vect{w}_q,\lambda) = &-\vect{w}_q^{\rm H}\left(\tilde{\vect{F}}_q + \frac{\tau}{2}\vect{I}_{KN} \right)\vect{w}_q + \Re\left\{(\vect{v}_q^t)^{\rm H}\vect{w}_q \right\} \\
		&- \lambda\left( \norm{\vect{w}_q}^2 - P_q \right),
	\end{aligned}
\end{equation}
where $\lambda \geq 0$ is the Lagrange multiplier associated with the transmit power constraint. Then, the optimal $\vect{w}_q$ can be derived by computing the derivative of $\mathcal{L}(\vect{w}_q,\lambda)$ with respect to $\vect{w}_q$ and then equating it with zero, which yields the following closed-form expression for each precoder:
\begin{equation} \label{eqn:optimal_w}
	\vect{w}_q^{\rm opt}(\lambda) = \frac{1}{2}\left(\tilde{\vect{F}}_q + \left(\frac{\tau}{2} + \lambda\right)\vect{I}_{KN} \right)^{-1} \vect{v}_q^t.
\end{equation}
Clearly, the optimal $\vect{w}_q$ depends on $\lambda$. To obtain $\lambda^{\rm opt}$, it suffices to consider Slater's condition \cite{Boyd_2004} according to which it should hold that:
\begin{equation}
	\lambda^{\rm opt}\left( (\vect{w}_q^{\rm opt})^{\rm H} \vect{w}_q^{\rm opt} - P_q \right) = 0.
\end{equation}
The above equation can be finally solved based on bisection search similarly to \cite[Corollary 1]{PLS2022_counteracting_all}.

\subsection{Local RIS Reflection Configuration Optimization} \label{Sec:RIS_Config}
According to the RIS element frequency response model in Section~\ref{Sec:Freq_Response}, the reflection configuration vector at each $q$-th BD RIS for each $k$-th SC (i.e., $\vect{\phi}_{q,k}$) depends on the parameters included in $\vect{c}_q$. It can thus be optimized by solving the following optimization sub-problem:
\begin{align*}
	\mathcal{OP}_3: \,\max_{\vect{c}_q} \, & \,\, - \frac{\tau}{2}\norm{\vect{c}_q - \vect{c}_q^t}^2 + \Re\{ (\vect{\gamma}_{\vect{c}_q}^t + \underline{\vect{\pi}}_q^t)^{\rm H}(\vect{c}_q - \vect{c}_q^t)\} \\
	\text{s.t.} & \quad C_{\min} \leq [\vect{c}_q]_m \leq C_{\max} \, \, \forall m=1,2,\ldots,M.
\end{align*}
Clearly, $\mathcal{OP}_3$ is a concave optimization problem since its objective function is the sum of a concave (i.e., the negative squared norm) and a linear term. To proceed with its solution, it suffices to obtain analytic expressions for $\vect{\gamma}_{\vect{c}_q}^t$ and $\underline{\vect{\pi}}_{\vect{c}_{q}}^t$, as presented in the next Theorem.

\begin{Thm} \label{thm:RIS_vectors}
The vectors $\vect{\gamma}_{\vect{c}_{q}}^t$ and $\underline{\vect{\pi}}_{\vect{c}_{q}}^t$ appearing in $\mathcal{OP}_3$ are given by the following analytic expressions: 
	\begin{align}
		&\begin{aligned}
			\mathllap{\vect{\gamma}_{\vect{c}_q}^t} =& \frac{2}{\ln(2)}\sum_{k=1}^K \Bigg(\frac{1}{(1+\operatorname{snr}_{qk}^t)\operatorname{MUI}_{qk}^t} \\
			&\times \Re\Bigg\{\diag\left\{\frac{\partial([\vect{\phi}_{q,k}]_1)^*}{\partial C_{1q}},\ldots,\frac{\partial([\vect{\phi}_{q,k}]_M)^*}{\partial C_{Mq}}\right\} \\ 
			&\times\operatorname{vec}_{\rm d}\left(\vect{A}_{qq,k}^{\rm H} + \vect{C}_{qq,k}^{\rm T}\vect{\Phi}_{q,k}^t\vect{B}_{qq,k}^{\rm T}\right)\Bigg\} \Bigg),
		\end{aligned} \label{eqn:gamma_RIS}\\
		&\begin{aligned}
			\mathllap{\underline{\vect{\pi}}_{\vect{c}_{q}}^t} =& -\frac{2}{\ln(2)}\sum_{j \neq q}^Q \sum_{k=1}^K \Bigg(\frac{\operatorname{snr}_{jk}^t}{(1+\operatorname{snr}_{jk}^t)\operatorname{MUI}_{jk}^t} \\
			&\times \Re\Bigg\{\diag\left\{\frac{\partial([\vect{\phi}_{q,k}]_1)^*}{\partial C_{1q}},\ldots,\frac{\partial([\vect{\phi}_{q,k}]_M)^*}{\partial C_{Mq}}\right\} \\ 
			&\times\operatorname{vec}_{\rm d}\left(\vect{A}_{qj,k}^{\rm H} + \vect{C}_{qq,k}^{\rm T}\vect{\Phi}_{q,k}^t\vect{B}_{qj,k}^{\rm T}\right)\Bigg\} \Bigg), \label{eqn:pricing_RIS}
		\end{aligned}
	\end{align}
	where the following matrix definitions have been used: $\vect{A}_{qq,k} \triangleq \vect{H}_{qq,k}\vect{w}_{qk}\vect{w}_{qk}^{\rm H}\vect{h}_{qq,k}\vect{g}_{qq,k}^{\rm H}\vect{S}_q$, $\vect{A}_{qj,k} \triangleq \vect{H}_{qq,k}\vect{w}_{qk}\vect{w}_{qk}^{\rm H}\vect{h}_{qj,k}\vect{g}_{qj,k}^{\rm H}\vect{S}_q$, $\vect{B}_{qq,k} \triangleq \vect{S}_q^{\rm T}\vect{g}_{qq,k}\vect{g}_{qq,k}^{\rm H}\vect{S}_q$, $\vect{B}_{qj,k} \triangleq \vect{S}_q^{\rm T}\vect{g}_{qj,k}\vect{g}_{qj,k}^{\rm H}\vect{S}_q$, and $\vect{C}_{qq,k} \triangleq \vect{H}_{qq,k}\vect{w}_{qk}\vect{w}_{qk}^{\rm H}\vect{H}_{qq,k}^{\rm H}$. In addition, the partial derivatives of $[\vect{\phi}_{q,k}]_m$ $\forall$$q,k,m$ with respect to each RIS tunable capacitance $C_{mq}$ can be computed as follows: 
	\begin{align} \label{eqn:der_C_m}  
		\nonumber\frac{\partial([\vect{\phi}_{q,k}]_m)^*}{\partial C_{mq}} =&\frac{-2}{\left(\mathcal{N}_{mq}^*(f_k,C_{mq}) + \mathcal{D}_{mq}^*(f_k,C_{mq}) \right)^2}\\
        &\times\Bigg( \frac{\partial \mathcal{N}_{mq}^*(f_k,C_{mq})}{\partial C_{mq}} \mathcal{D}_{mq}^*(f_k,C_{mq})\\
        &\hspace{0.68cm}- \mathcal{N}_{mq}^*(f_k,C_{mq}) \frac{\partial \mathcal{D}_{mq}^*(f_k,C_{mq})}{\partial C_{mq}} \Bigg),\nonumber
	\end{align}
	where respectively following \eqref{eqn:numerator_s} and \eqref{eqn:denominator_s} holds that: 
	\begin{align}
		\frac{\partial \mathcal{N}_{mq}^*(f_k,C_{mq})}{\partial C_{mq}} &= -(\kappa f_k)^2(L_1 + L_2) - \jmath \kappa f_k R, \\
		\frac{\partial \mathcal{D}_{mq}^*(f_k,C_{mq})}{\partial C_{mq}} &= -\jmath \kappa f_k \frac{L_1}{Z_0}(-(\kappa f_k)^2 L_2 - \jmath \kappa f_k R).
	\end{align}
\end{Thm}

\begin{IEEEproof}
	The detailed proof is delegated in Appendix \ref{appx:RIS_vectors}.    
\end{IEEEproof}

Based on the previously derived $\vect{\gamma}_{\vect{c}_q}^t$ and $\underline{\vect{\pi}}_{\vect{c}_{q}}^t$ expressions, $\mathcal{OP}_3$ can be solved in closed form, as presented in the following Corollary.
\begin{Cor} \label{cor:RIS_cl_form_solution}
	Let $\vect{\beta}_q \triangleq \tau\vect{c}_q^t + \vect{\gamma}_{\vect{c}_q}^t + \underline{\vect{\pi}}_{\vect{c}_{q}}^t$. Then, the optimal solution to $\mathcal{OP}_3$ is given in closed form as follows:
	\begin{equation} \label{eqn:RIS_solution}
		[\vect{c}_q]_m^{\rm opt} =  \begin{cases}
			C_{\min}, & \text{if}\,\,\frac{1}{\tau}[\vect{\beta}_q]_m < C_{\min} \\
			C_{\max}, & \text{if}\,\,\frac{1}{\tau}[\vect{\beta}_q]_m >C_{\max} \\
			\frac{1}{\tau}[\vect{\beta}_q]_m, & \text{otherwise}
		\end{cases}.
	\end{equation}
\end{Cor}

\begin{IEEEproof}
	The optimal solution to $\mathcal{OP}_3$ can be found by expanding the norm as well as the linear term included in its objective. In fact, after some algebraic manipulations, the objective of $\mathcal{OP}_3$ reduces to $-\frac{\tau}{2}\vect{c}_q^{\rm T}\vect{c}_q + \vect{\beta}_q^{\rm T}\vect{c}_q$, where the constant terms have been omitted. Next, by noticing that the latter expression is concave with respect to $\vect{c}_q$ and that each RIS element's response is independent from the others, \eqref{eqn:RIS_solution} can be derived by the first-order condition of optimality, taking into account the box constraints; this completes the proof.
\end{IEEEproof}

\subsection{Local RIS Switch Selection Matrix Optimization} \label{Sec:Non_Diag_Design}
The design of the switch selection matrix $\vect{S}_q$ at each $q$-th BD RIS, which is restricted to the set $\mathcal{S}$, reduces to the following simplified optimization sub-problem:
\begin{align*}
	\mathcal{OP}_4: \max_{\vect{S}_q \in \mathcal{S}} \, & \,\, -\frac{\tau}{2}\norm{ \vect{S}_q - \vect{S}_q^t}_{\rm F}^2 + <\vect{\Gamma}_{\vect{S}_q}^t + \vect{\Pi}_{\vect{S}_q}^t,\vect{S}_q - \vect{S}_q^t>,
\end{align*}
whose solution needs to derive, at first, the matrices $\vect{\Gamma}_{\vect{S}_q}^t$ and $\vect{\Pi}_{\vect{S}_q}^t$, as presented in the sequel.
\begin{Cor} \label{thm:Sel_Mat_vectors}
	Let the following matrix definitions:
	\begin{align} 
		\vect{F}_{qq,k} &\triangleq \vect{\Phi}_{q,k}\vect{H}_{qq,k}\vect{w}_{qk}\vect{w}_{qk}^{\rm H}\vect{h}_{qq,k}\vect{g}_{qq,k}^{\rm H}, \label{eqn:S_q_matrices_1} \\
		\vect{F}_{qj,k} &\triangleq \vect{\Phi}_{q,k}\vect{H}_{qq,k}\vect{w}_{qk}\vect{w}_{qk}^{\rm H}\vect{h}_{qj,k}\vect{g}_{qj,k}^{\rm H}, \label{eqn:S_q_matrices_2} \\
		\vect{K}_{qq,k} &\triangleq \vect{\Phi}_{q,k}\vect{H}_{qq,k}\vect{w}_{qk}\vect{w}_{qk}^{\rm H}\vect{H}_{qq,k}^{\rm H}\vect{\Phi}_{q,k}^{\rm H}, \label{eqn:S_q_matrices_3} \\
		\vect{G}_{qq,k} &\triangleq \vect{g}_{qq,k}\vect{g}_{qq,k}^{\rm H}, \,\,\vect{G}_{qj,k} \triangleq \vect{g}_{qj,k}\vect{g}_{qj,k}^{\rm H} \label{eqn:S_q_matrices_4}.
	\end{align}
	Then, it holds that $\vect{\Gamma}_{\vect{S}_q}^t$ and $\vect{\Pi}_{\vect{S}_q}^t$ can be expressed as follows:
	\begin{align}
		&\begin{aligned}
			\mathllap{\vect{\Gamma}_{\vect{S}_q}^t} = \frac{2}{\ln(2)}&\sum_{k=1}^K \frac{1}{(1+\operatorname{snr}_{qk}^t)\operatorname{MUI}_{qk}^t}\\
			&\times\left(\vect{F}_{qq,k} + \vect{K}_{qq,k}(\vect{S}_q^t)^{\rm T}\vect{G}_{qq,k} \right)^{\rm T}, \\
		\end{aligned}  \label{eqn:gamma_Sel_Mat}\\
		&\begin{aligned} 
			\mathllap{\vect{\Pi}_{\vect{S}_q}^t} = -\frac{2}{\ln(2)}&\sum_{j\neq q}^Q\sum_{k=1}^K\frac{\operatorname{snr}_{jk}^t}{(1+\operatorname{snr}_{jk}^t)\operatorname{MUI}_{jk}^t}\\
			&\times\left(\vect{F}_{qj,k} + \vect{K}_{qq,k}(\vect{S}_q^t)^{\rm T}\vect{G}_{qj,k} \right)^{\rm T}.
		\end{aligned} \label{eqn:pricing_Sel_Mat}
	\end{align}
\end{Cor}
\begin{IEEEproof}
	See Appendix \ref{appx:Sel_Mat_vectors}.
\end{IEEEproof}

By expanding the norm in the objective function of $\mathcal{OP}_4$ and omitting the constant terms, this problem can be equivalently re-written as follows:
\begin{align*}
	\mathcal{OP}_4: \max_{\vect{S}_q} \, & \,\, \trace\left( \Re\left\{\vect{\Gamma}_{\vect{S}_q}^t + \vect{\Pi}_{\vect{S}_q}^t + \tau\vect{S}_q^t\right\}^{\rm H}\vect{S}_q \right) \\
	\text{s.t.} & \quad [\vect{S}_q]_{i,j}\in\{0,1\},\forall i,j = 1,2,\ldots,M, \\
	& \quad \sum_{i=1}^M[\vect{S}_q]_{i,j} = 1, \forall j = 1,2,\ldots,M, \\
	& \quad \sum_{j=1}^M[\vect{S}_q]_{i,j} = 1, \forall i = 1,2,\ldots,M,
\end{align*}
where the quadratic term $\trace(\vect{S}_q\vect{S}_q^{\rm T})$ has been treated as a constant equal to $M$, due to the fact that $\vect{S}_q$ is, by definition, a permutation matrix. $\mathcal{OP}_4$ is a combinatorial problem that belongs to the class of Linear Sum Assignment Problems (LSAPs). Thus, according to \cite[Theorem 4.2]{burkard2012assignment}, its optimal solution can be obtained by relaxing the set of binary constraints, in particular, by letting $[\vect{S}_q]_{i,j} \in [0,1] \forall i,j$, since the rest of the constraints formulate a totally unimodular constraint matrix. Then, the relaxed problem can be efficiently solved either by linear programming techniques or other strategies, such as computing permutations of sparse matrices \cite{duff2001algorithms} (especially when $M$ is large enough).

\subsection{Overview of the Proposed Distributed Solution} \label{Sec:OP_Overall_Algorithm}
\begin{algorithm}[!t]
	\begin{algorithmic}[1]
		\caption{Proposed D-SCA Design Solving $\mathcal{OP}$}
		\label{alg:OP_Overall_Distributed_Algorithm}
		\State \textbf{Input:} $t=0$, $\{\alpha^t\}\geq 0$, $\tau>0$, $\epsilon > 0$, $Q$, as well as feasible $\widetilde{\vect{w}}^{(0)}$, $\widetilde{\vect{c}}^{(0)}$, $\widetilde{\vect{S}}^{(0)}$, and $\overline{\mathcal{R}}^{(0)}$ as defined in \eqref{eqn:total_rate}.
		\State Compute $\vect{\phi}_{q,k}^{(0)}\,\,\forall q,k$ as a function of $\widetilde{\vect{c}}^{(0)}$ using \eqref{eqn:RIS_freq_response}.
		\For{$ t = 1,2,\ldots$}
		\For{$ q = 1,2,\ldots,Q$}
		\State 
		\parbox[t]{\dimexpr\linewidth-\algorithmicindent}{%
			Compute $\vect{f}_{qq,k}$ and $\vect{f}_{jq,k}\,\,\forall j\neq q$ according to \eqref{eqn:total_channels_1} 
			
			and \eqref{eqn:total_channels_2}.
		}
		\State Compute the pricing vector $\overline{\vect{\pi}}_{qk}^t$ using \eqref{eqn:pricing_w_q}.
		\State Compute $a_{qk}^t$ using \eqref{eqn:surrog_a} and $\vect{b}_{qk}^t$ using \eqref{eqn:surrog_b}.
		\State 
		\parbox[t]{\dimexpr\linewidth-\algorithmicindent}{%
			Formulate the block diagonal matrix $\tilde{\vect{F}}_q$ and 
			
			compute the vectors $\tilde{\vect{f}}_q = [(\vect{b}_{q1}^t)^{\rm T},\ldots,(\vect{b}_{qK}^t)^{\rm T}]^{\rm T}$
			
			and $\vect{v}_q^t = \overline{\vect{\pi}}_q^t + 2\tilde{\vect{f}}_q + \tau \vect{w}_q^{t-1}$.
		}
		\State 
		\parbox[t]{\dimexpr\linewidth-\algorithmicindent}{%
			Compute $\vect{w}_q^t$ according to \eqref{eqn:optimal_w} and a bisection 
			
			method.
		}
		\State Compute $\vect{\gamma}_{\vect{c}_q}^t$ and $\underline{\vect{\pi}}_{\vect{c}_q}^t$ according to Theorem~\ref{thm:RIS_vectors}.
		\State Compute $\vect{c}_q^t$ according to Corollary~\ref{cor:RIS_cl_form_solution}.
		\State Compute $\vect{\Gamma}_{\vect{S}_q}^t$ and $\vect{\Pi}_{\vect{S}_q}^t$ according to Corollary~\ref{thm:Sel_Mat_vectors}.
        \State \parbox[t]{\dimexpr\linewidth-\algorithmicindent}{%
        Collect $\vect{\Pi}_q^t = \{\overline{\vect{\pi}}_{qk}^t,\underline{\vect{\pi}}_{\vect{c}_q}^t,\vect{\Pi}_{\vect{S}_q}^t\}$ and transmit to 
        
        users $j\neq q$.
        }
		\State 
		\parbox[t]{\dimexpr\linewidth-\algorithmicindent}{%
			Solve the LSAP $\mathcal{OP}_4$ numerically to compute $\vect{S}_q^t$.
		}
		\State Obtain $\widehat{\vect{X}}_q^t = \left\{\widehat{\vect{w}}_q^t,\widehat{\vect{c}}_q^t,\widehat{\vect{S}}_q^t\right\}$ and $\vect{X}_q^{t+1}$ using \eqref{eqn:OP1_ascent_solution}.
		\EndFor
		\If $\left\lvert\left(\overline{\mathcal{R}}^{(t)} - \overline{\mathcal{R}}^{(t-1)}\right)/\overline{\mathcal{R}}^{(t)}\right\rvert \leq \epsilon$, \textbf{break}; 
		\EndIf
		\EndFor
		\State \textbf{Output:} $\widetilde{\vect{w}}^{(t)}$, $\widetilde{\vect{c}}^{(t)}$, and $\widetilde{\vect{S}}^{(t)}$.
	\end{algorithmic}
\end{algorithm}

According to the previous subsections, the best-response mapping $\widehat{\vect{X}}_q^t$ can be computed by solving $\mathcal{OP}_2$, $\mathcal{OP}_3$, and $\mathcal{OP}_4$. The solution to $\mathcal{OP}$ for the precoding vectors at the $Q$ BSs and the tunable capacitances as well as the switch selection matrices for the $Q$ BD RISs is then computed for each algorithmic iteration $t+1$ and $\forall$$q = 1,2,\ldots,Q$ as follows:
\begin{equation} \label{eqn:OP1_ascent_solution}
	\vect{X}_q^{t+1} = \vect{X}_q^t + \alpha^t\left( \widehat{\vect{X}}_q^t - \vect{X}_q^t \right),
\end{equation}
where $\alpha^t$ represents the possibly time-varying step size. This step-size must be chosen either as constant and sufficiently small or having diminishing values, according to classical stochastic approximation rules, to guarantee convergence of the overall iterative procedure to a local maximum of $\mathcal{OP}$~\cite{scutari2013decomposition_all} (examples are given in Section~\ref{Sec:Numerical} that follows). The proposed distributed solution to $\mathcal{OP}$, termed as Distributed Successive Concave Approximation (D-SCA), is summarized in Algorithm~\ref{alg:OP_Overall_Distributed_Algorithm}. Interestingly, $\mathcal{OP}$ (i.e., Steps $4$ to $16$ in Algorithm~\ref{alg:OP_Overall_Distributed_Algorithm}) can be solved distributively by each user $q$ (e.g., by each $q$-th BS), once the MUI is estimated locally at each $q$-th UE (i.e., the term $\sigma_{q}^2+ \sum_{j=1}^K \lvert \vect{f}_{jq,k}^{\rm H} \vect{w}_{jk}\rvert^2 $ in \eqref{eqn:first_surrogate} $\forall$$k,q$), and the price vectors $\vect{\Pi}_q^t$ in (\ref{eqn:pricing_def}) are transmitted to user $q$ by each $j$-th user with $j\neq q$. 

To run Algorithm~\ref{alg:OP_Overall_Distributed_Algorithm}, feasible points for $\widetilde{\vect{w}}^{(0)}$, $\widetilde{\vect{c}}^{(0)}$, and $\widetilde{\vect{S}}^{(0)}$ are needed. However, these parameters are coupled with each other, hence, these points become quite difficult to obtain directly. To handle this limitation, we commence by randomly initializing $\widetilde{\vect{c}}^{(0)}$ and $\widetilde{\vect{S}}^{(0)}$ such that they satisfy the respective constraints. Then, we compute $\vect{f}_{qq,k}$ for each $q$-th user at each $k$-th SC via \eqref{eqn:total_channels_1}, and next realize the typical maximum ratio transmission precoding vector as $\bar{\vect{w}}_{qk}^{(0)} = \frac{1}{\|\vect{f}_{qq,k}\|}\vect{f}_{qq,k}$. To obtain a feasible precoder, we set $\vect{w}_{qk}^{(0)} = \sqrt{\frac{P_q}{\sum_{k=1}^K \|\bar{\vect{w}}_{qk}^{(0)}\|^2}}\bar{\vect{w}}_{qk}^{(0)}$ $\forall q$ and finally construct $\widetilde{\vect{w}}^{(0)}$ by stacking all vectors per SC and per user. 
Following this randomized initialization scheme, we have observed through numerous simulations that it has minor impact on the overall performance, thus, avoiding instabilities for Algorithm~\ref{alg:OP_Overall_Distributed_Algorithm}.

\subsection{Convergence and Complexity Analysis} \label{Sec:Convergence_Complexity_Analysis}
The convergence properties of the proposed Algorithm~\ref{alg:OP_Overall_Distributed_Algorithm}, which provides our distributed design for the considered multi-RIS-empowered multiple access system, are described in the following theorem.

\begin{Thm} \label{thm:KKT_Proof}
    Algorithm~\ref{alg:OP_Overall_Distributed_Algorithm} either converges in a finite number of iterations to a stationary solution of $\mathcal{OP}$ or every limit point of the sequence $\{\vect{X}^t\}_{t = 0}^{\infty}$ is a Karush-Kuhn-Tucker (KKT) point of $\mathcal{OP}$. Both of such points are local maxima of \eqref{eqn:total_rate}. 
\end{Thm}

\begin{IEEEproof}
    The proof is delegated in Appendix~\ref{appx:KKT_Proof}.
\end{IEEEproof}

The computational complexity of Algorithm~\ref{alg:OP_Overall_Distributed_Algorithm} is analyzed via inspection of its main algorithmic steps, as follows. In Step $5$, the computation of the cascaded channels involves the multiplication of (excluding subscripts for clarity) the channel vector $\vect{g}$, the permutation matrix $\vect{S}$, the diagonal matrix $\vect{\Phi}$ and the channel matrix $\vect{H}$, which requires $\mathcal{O}(M(N+1))$ computations. The calculation of the pricing vector in Step $6$ results in $\mathcal{O}(N)$ complexity, since it requires the computation of an inner product; the same holds for Step $7$. Step $8$ has negligible computational cost, while Step $9$ involves the inversion of a $KN\times KN$ square matrix, which yields $\mathcal{O}((KN)^3)$ computational complexity. Note that the computation of the optimal Lagrange multiplier via the bisection method is negligible. In Step $10$, the worst case complexity is $\mathcal{O}(M)$ due to the multiplication of a diagonal matrix with a vector, whereas the cost for Step $11$ is negligible. In addition, Step $12$ requires the worst case complexity of $\mathcal{O}(M^2)$, since it involves the inner product operation between matrices. In Step $14$, the worst case computational complexity for the LSAP solution is $\mathcal{O}(M(\mu + M\log M))$ \cite{burkard2012assignment}, where $\mu$ is a constant related to the number of admitted $(i,j)$ assignments, equal to the cardinality of the corresponding edges (for more details the reader is referred to \cite[Chapter 4]{burkard2012assignment}). Putting all above together, the total complexity of Algorithm~\ref{alg:OP_Overall_Distributed_Algorithm} for solving $\mathcal{OP}$ in a distributed manner is given by:
\begin{equation} \label{eqn:OP_total_complexity}
	\begin{aligned}
		C_{\mathcal{OP}} = \mathcal{O}\Bigg(&T_{\max} Q \max\Big\{ M(N+1),(KN)^3,M^2,\\
		&M(\mu + M\log M)\Big\} \Bigg),
	\end{aligned}    
\end{equation}
where $T_{\max}$ represents the total number of iterations until convergence. 

In terms of the information exchange overhead, our distributed scheme necessitates that each $q$-th BS computes the pricing vectors $\vect{\Pi}_q^t=\{\overline{\boldsymbol{\pi}}_{qk}^t,\underline{\boldsymbol{\pi}}_{\vect{c}_q}^t,\vect{\Pi}_{\vect{S}_q}^t\}$ (or matrices for the case of $\vect{S}_q$), and then transmits them to each $j$-th BS with $j\neq q$. Recall that the dimension of vector $\overline{\boldsymbol{\pi}}_{qk}^t$ is $KN\times 1$, while the corresponding vector for each $q$-th RIS $\underline{\boldsymbol{\pi}}_{\vect{c}_q}^t$ has the dimension $M\times 1$. In addition, the size of the pricing matrix $\vect{\Pi}_{\vect{S}_q}^t$ is $M\times M$. Overall, for the considered multi-RIS-empowered interference channel with $Q$ users, the BS cooperation overhead is equal to $Q(KN + M(M + 1))$.

\section{Numerical Results and Discussion} \label{Sec:Numerical}
In this section, we include detailed simulation results for the performance of the proposed distributed design presented in Section~\ref{Sec:Design_SCA}, which were obtained through the numerical evaluation of the achievable sum-rate performance expression in~\eqref{eqn:total_rate}, including the pre-sum multiplicative factor $\frac{1}{K}$. We have particularly implemented Algorithm~\ref{alg:OP_Overall_Distributed_Algorithm} to design the linear precoding vectors at the BSs as well as the tunable capacitances and the switch selection matrices at the BD RIS with the proposed distributed manner.

\subsection{Simulation Setup} \label{Sec:Sims_Setup}
In our simulations, all nodes were considered positioned on the $3$-Dimensional ($3$D) Cartesian coordinate system with coordinates given by the triad $(x,y,h)$, where $x$ and $y$ denote the coordinates on the $x$- and $y$-axis, respectively, while $h$ represents the node's height, i.e., its positive value on the $z$-axis. We have considered $Q = 4$ BSs located in a square of width $w = 60\,m$, and placed BS$_1$ at the origin in the height $h_{\rm BS_1}$, BS$_2$ at the location $(w,0,h_{\rm BS_2})$, BS$_3$ at $(0,w,h_{\rm BS_3})$, and BS$_4$ at $(w,w,h_{\rm BS_4})$ with $h_{\rm BS_q} = 5\,m$ $\forall q=1,2,3$, and $4$. Accordingly, the four UEs share the same coordinates $y_{\rm UE} = 60\,m$ and $h_{\rm UE} = 1.5\,m$ on the $yz$-plane, while differing in the $x$-coordinate as follows: $x_{\rm UE_1} = 27$, $x_{\rm UE_2} = 33$, $x_{\rm UE_3} = 28.5$, and $x_{\rm UE_4} = 31.5\,m$. In addition, the coordinates of the four BD RISs on the $xy$-plane were fixed to $(27.75,62.5)$ for RIS$_1$, $(32.25,62.5)$ for RIS$_2$, $(27.75,57.5)$ for RIS$_3$, and RIS$_4$ at $(32.25,57.5)$, while all shared the same $z$-coordinate value $h_{\rm RIS} = 3\,m$. All wireless channels were modeled as wideband fading channels with $D$ delay taps in their time-domain impulse responses, whose elements were assumed to follow the circularly symmetric complex Gaussian distribution. To construct each channel at the frequency domain, we first considered the block cyclic matrices $\widetilde{\vect{H}}_{qq} \in \mathbb{C}^{K\times KN}$, $\widehat{\vect{H}}_{qq} \in \mathbb{C}^{MK\times KN}$ and $\widetilde{\vect{G}}_{qq} \in \mathbb{C}^{K\times KM}$ for each $q$-th user, whose first blocks were defined as $\left[ (\widetilde{\vect{h}}_{qq,0})^{\rm H},\dots,(\widetilde{\vect{h}}_{qq,D-1})^{\rm H},\vect{0}_{N}^{\rm T},\ldots,\vect{0}_{N}^{\rm T} \right]^{\rm H}$, $\left[ (\widehat{\vect{H}}_{qq,0})^{\rm H},\dots,(\widehat{\vect{H}}_{qq,D-1})^{\rm H},\vect{0}_{N\times M}^{\rm T},\ldots,\vect{0}_{N \times M}^{\rm T} \right]^{\rm H}$ and $\left[ (\widetilde{\vect{g}}_{qq,0})^{\rm H},\dots,(\widetilde{\vect{g}}_{qq,D-1})^{\rm H},\vect{0}_{M}^{\rm T},\ldots,\vect{0}_{M}^{\rm T} \right]^{\rm H}$, respectively. In the latter expressions, $\widetilde{\vect{h}}_{qq,d}$, $\widehat{\vect{H}}_{qq,d}$, and $\widetilde{\vect{g}}_{qq,d}$ represent the respective channel's impulse response value at the $d$-th delay tap ($d = 0,1,\dots,D-1$). Consequently, the latter matrices can be arranged as sequences of cyclic matrices to apply the Discrete Fourier Transform (DFT) (via the normalized DFT matrix $\vect{F}_{\rm DFT}$) and express them in the frequency domain as described in~\cite{bjornson2024introduction,li2021_wideband_practical_all}. We have considered distance-dependent pathloss between any two nodes $i$ and $j$ with distance $d_{i,j}$ ($i$ and $j$ take values from the string set $\{\rm BS,UE,RIS\}$), which was modeled as ${\rm PL}_{i,j} = {\rm PL_0}(d_{i,j}/d_0)^{-\alpha_{i,j}}$ with ${\rm PL_0} = (\frac{\lambda_c}{4\pi})^2$ denoting the signal attenuation at the reference distance $d_0 = 1$ $m$ and $\lambda_c$ represents the carrier signal's wavelength. The distance $d_{i,j}$ was computed per $q$-th BS's antenna element and $q$-th UE, $q$-th BS's antenna element and $q$-th RIS's element, and $q$-th RIS's element and $q$-th UE, to obtain the pathloss of the channels $\vect{h}_{qq,k}$, $\vect{H}_{qq,k}$ and $\vect{g}_{qq,k}$, respectively. For the pathloss exponents, we have set $\alpha_{\rm BS,UE} = 3.7$, $\alpha_{\rm BS,RIS} = 2.6$, and $\alpha_{\rm RIS,UE} = 2.2$. We further assumed a uniform linear array and a uniform planar array deployed on the $xz$-plane for each BS and each BD RIS, respectively, both with element spacing equal to $\lambda_c/2$.

In the performance results that follow, we have set equal transmit power budget at all BSs and noise variances at all UEs, in particular, $P_q = P$ and $\sigma_q^2=\sigma^2=-90$ dBm $\forall q=1,2,3$, and $4$. The carrier frequency was set as $f_c = 3.5$ GHz, the bandwidth as ${\rm BW} = 100$ MHz, the number of SCs as $K = 64$ with the central frequency of each $k$-th SC defined as $f_k \triangleq f_c + (k - \frac{K + 1}{2})\frac{{\rm BW}}{K}$ $\forall k = 1,2,\ldots,K$, the delay taps as $D = 16$, and the cyclic prefix length as $N_{\rm cp} = 16$. The circuital elements for each BD RIS were set as follows~\cite{zhu2013active_impedance_all,abeywickrama2020intelligent_all}: $L_1 = 2.5$ nH, $L_2 = 0.7$ nH, $R = 1$ ${\rm \Omega}$, the free space impedance as $\mathcal{Z}_0 = 377$ $\Omega$, $C_{\min} = 0.47$ pF, and $C_{\max} = 2.35$ pF. It is noted that the latter values correspond to practical RISs with varactor diodes \cite{SMV1231}, taking into account the loss resistance effect. For the parameters to run the proposed Algorithm~\ref{alg:OP_Overall_Distributed_Algorithm}, we have set the regularization parameter as $\tau = 1.85$ and the convergence threshold as $\epsilon = 10^{-3}$. For the step size, we have used the time varying rule $\alpha^t = \frac{\alpha^{t-1}+ a(t)}{1+b(t)}$ for each $(t+1)$-th algorithmic iteration as well as $\alpha^0 = 1$~\cite{scutari2013decomposition_all} for the updates with respect to the linear BS precoders $\vect{w}_q$'s and the BD RIS parameters $\vect{c}_q$'s, whereas, for the selection matrices $\vect{S}_q$'s, we have set $\alpha^t = \alpha = 1$ $\forall$$t$ to avoid violating the relevant variable constraints due to the update rule in \eqref{eqn:OP1_ascent_solution}. For all performance evaluation results that follow, we have used $200$ independent Monte Carlo realizations.

\subsection{Performance Evaluation} \label{Sec:Sims_Performance}
\begin{figure}[!t]
	\centering
	\includegraphics[width=3.45in]{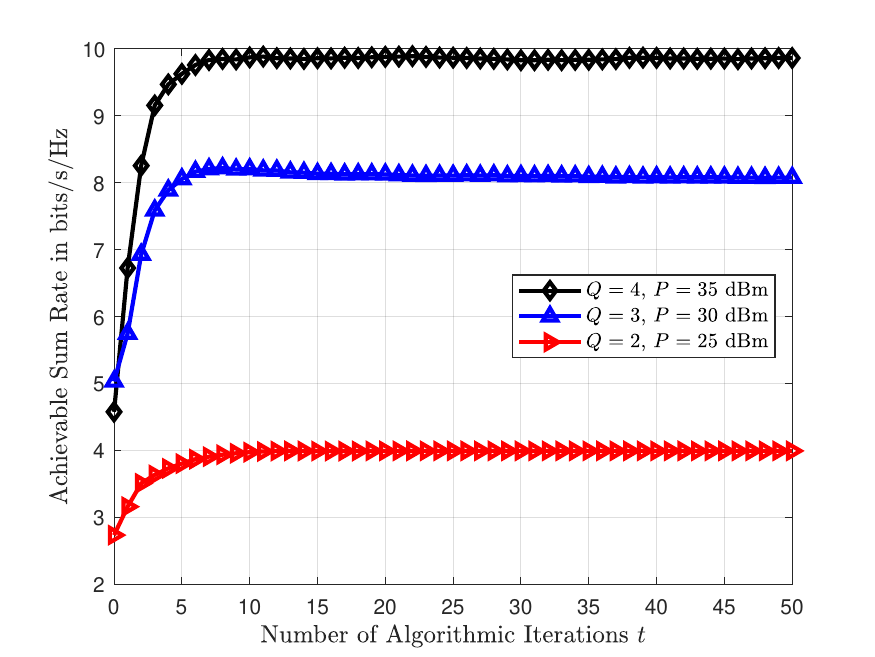}
	\caption{\small{Convergence of the achievable sum rate with the proposed Algorithm~\ref{alg:OP_Overall_Distributed_Algorithm} for different numbers of BSs $Q$ each with $N=4$ antenna elements, and different transmit power levels $P$, considering $Q$ BD RISs each with $M = 100$ elements and $Q$ single-antenna UEs}.}
	\label{fig:Rates_vs_Iter}
\end{figure}

The convergence of the proposed distributed design summarized in Algorithm~\ref{alg:OP_Overall_Distributed_Algorithm} is illustrated in Fig.~\ref{fig:Rates_vs_Iter} for different combinations of the total number of users $Q$ and transmit power budget values $P$. It can be observed that convergence is achieved within $T_{\max} = 10$ iterations for $Q=2$ and $P=25$~dBm, while slightly more iterations are required when $Q \geq 3$ and $P \geq 30$ dBm. For these cases, convergence is achieved with $T_{\max} = 20$ iterations. This behavior indicates that, as $Q$ increases, the proposed design is able to converge within limited number of algorithmic iterations, keeping $T_{\max}$ relatively small (cf. \eqref{eqn:OP_total_complexity}).

\begin{figure}[!t]
	\centering
	\includegraphics[width=3.45in]{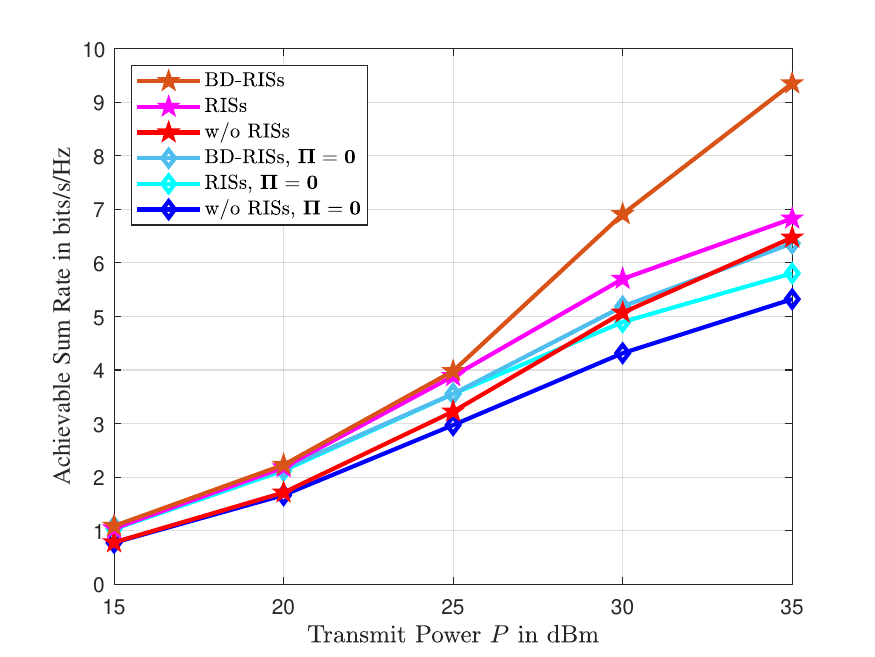}
	\caption{\small{Achievable sum-rate performance versus the transmit power $P$ of each BS for $Q=2$ users, considering BSs each with $N=4$ antennas and RISs each with $M=100$ elements. Cooperative and non-cooperative schemes with diagonal and BD RISs are compared.}}
	\label{fig:Rates_vs_TX_Power}
\end{figure}

We next compare the achievable sum-rate performance of the proposed distributed design, abbreviated in the results that follow as ``BD-RISs,'' with the following schemes: \textit{i}) ``w/o RISs,'' where no RISs are deployed and the system design is solely based on the direct link, i.e., only the BSs' precoders are optimized; \textit{ii}) ``RISs,'' where all deployed metasurfaces have a diagonal structure, and thus, $\vect{S}_q=\vect{I}_M$ for each $q$-th RIS. We have also evaluated three more special cases for which the pricing vectors/matrices were fixed to zero, either for the BD RISs or not, as well as for the case where no RISs are deployed. Based on the proposed design presented in Section~\ref{Sec:Design_SCA}, the latter cases correspond to the absence of cooperation among the nodes, and are next indicated with $\vect{\Pi} = \vect{0}$. For these cases, we have set the regularization parameter in Algorithm~\ref{alg:OP_Overall_Distributed_Algorithm} as $\tau = 1.25$. In Fig.~\ref{fig:Rates_vs_TX_Power}, we depict the achievable sum rates for all considered schemes as a function of the $P$ value for each BS, considering $Q=2$ users with $4$-antenna BSs and $100$-elements RISs. As illustrated, for all schemes, the performance follows a non-decreasing trend for increasing values of $P$. However, for $P \in [30,35]$ dBm, all rates except the proposed one (i.e., ``BD-RISs'') witness a smaller gain than that observed for $P \leq 30$ dBm. This behavior is attributed to the presence of more severe interference in the high Signal-to-Noise Ratio (SNR) regime. It is also shown that all schemes relying on BD RISs outperform their diagonal RIS counterparts; this is evident either in the cases with cooperation or not. With the proposed ``BD-RISs'' scheme necessitating user cooperation for pricing vectors/matrices exchange, the performance gain is approximately $20\%$ at $P=30$ dBm and $35\%$ when $P = 35$ dBm, verifying that BD RISs provide more degrees of freedom for optimization in comparison to diagonal RISs. At lower SNR values, the corresponding gain becomes smaller. Interestingly, when cooperation occurs, the ``w/o RISs'' scheme (i.e., only cooperative beamforming) slightly outperforms the ``BD-RISs, $\vect{\Pi} = \vect{0}$'' scheme, signifying the benefits of cooperation.

\begin{figure}[!t]
	\centering
	\includegraphics[width=3.45in]{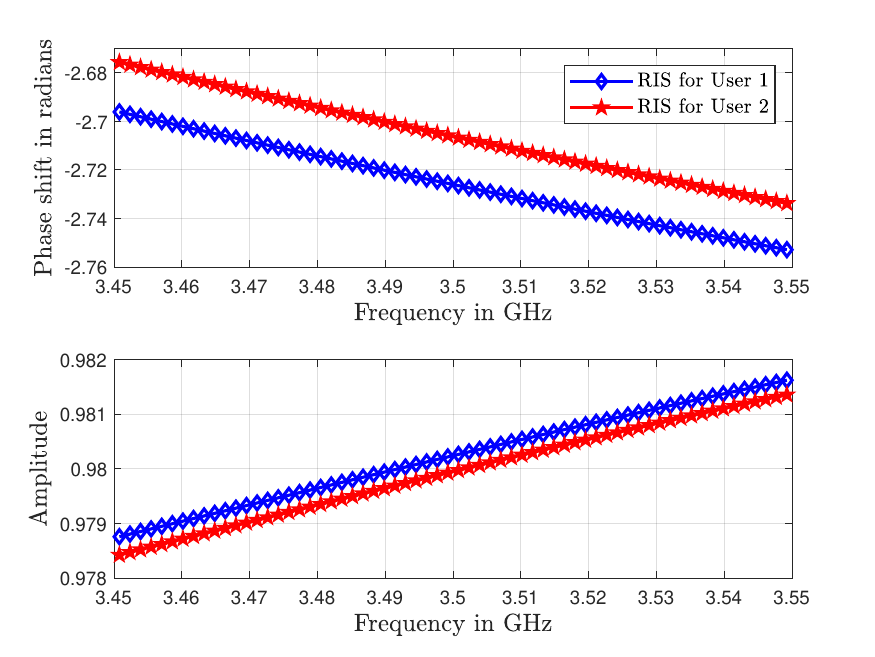}
	\caption{\small{The phase shift and amplitude values of the BD RIS reflection coefficient with the proposed ``BD-RISs'' scheme in Fig.~\ref{fig:Rates_vs_TX_Power} averaged over the considered independent channel realizations, RIS elements, and transmit power levels, with respect to the frequency SC for the considered bandwidth of ${\rm BW} = 100$ MHz.}}
	\label{fig:Ampl_Phase_freq}
\end{figure}
The amplitude and phase values of the elements of the two optimized BD RISs with the proposed ``BD-RISs'' scheme in Fig.~\ref{fig:Rates_vs_TX_Power} are plotted for all considered SCs in Fig.~\ref{fig:Ampl_Phase_freq}. The depicted values have been averaged over all independent Monte Carlo realizations, all $M=100$ RIS elements, and all transmit power $P$ levels in $[15,35]$ dBm. It is evident that both the amplitudes and phases of the reflection coefficients are non-constant functions with respect to the frequency, showcasing the importance of considering wideband models for the BD RISs when dealing with OFDM systems. This dependency is more pronounced for the phase shifts (their range of values lies in the interval $[-2.7,-2.65]$ radians), while for the amplitude values, it can be observed from the bottom sub-figure in Fig.~\ref{fig:Ampl_Phase_freq} that the changes are smaller following an increasing pattern as the frequency increases.

\begin{figure}[!t]
	\centering
	\includegraphics[width=3.45in]{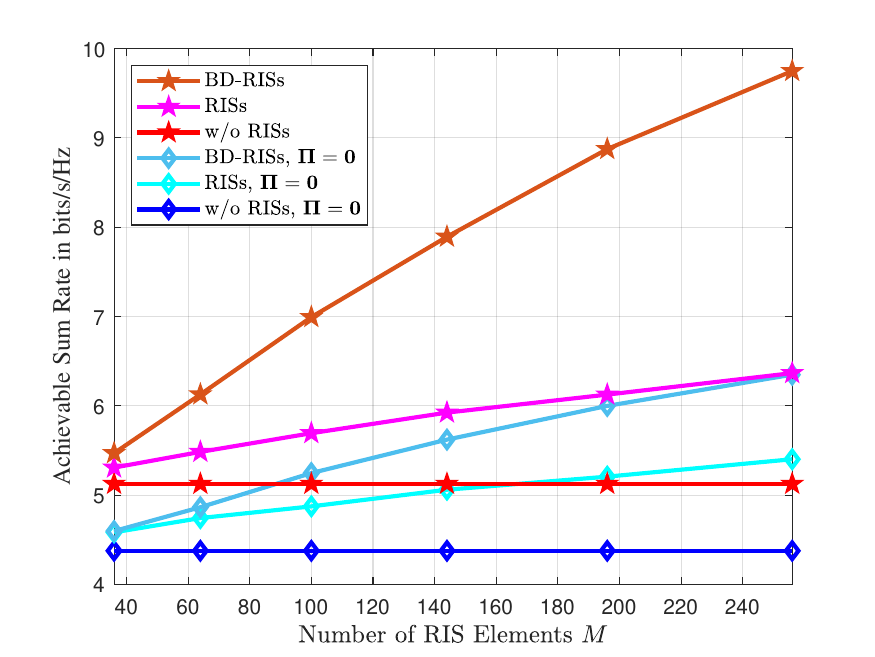}
	\caption{\small{Achievable sum-rate performance versus the common number $M$ of elements at each RIS for all investigated schemes, considering the transmit power $P=30$ dBm and $Q=2$ users with $N=4$ antenna elements at each BS.}}
	\label{fig:Rates_vs_M}
\end{figure}

The impact of the varying size of the RISs' on the achievable sum rate is illustrated in Fig.~\ref{fig:Rates_vs_M} for $P=30$ dBm. As expected, the number of RIS elements plays an important role in this metric for all investigated RIS-based schemes; all these schemes exhibit a non-decreasing trend as $M$ increases. This trend is, however, more pronounced with BD RISs instead of diagonal ones. Specifically, it can be seen that schemes with optimized diagonal RISs reach saturation points in the sum rate in contrast to BD RISs that yield constantly increasing performance. It is also shown that the gains of the RIS-based schemes are very small with respect to the ``w/o RISs'' cases for $M\leq 100$, while for this value and above, their performance increases significantly. Moreover, when the users cooperate with each other, it becomes apparent that the advantages of optimizing BD RISs are much more evident (no saturation point appears across the investigated range of $M$ values in the figure), resulting in a sum-rate improvement of more than $50\%$ when compared to diagonal RISs.

\begin{figure}[!t]
	\centering
	\includegraphics[width=3.45in]{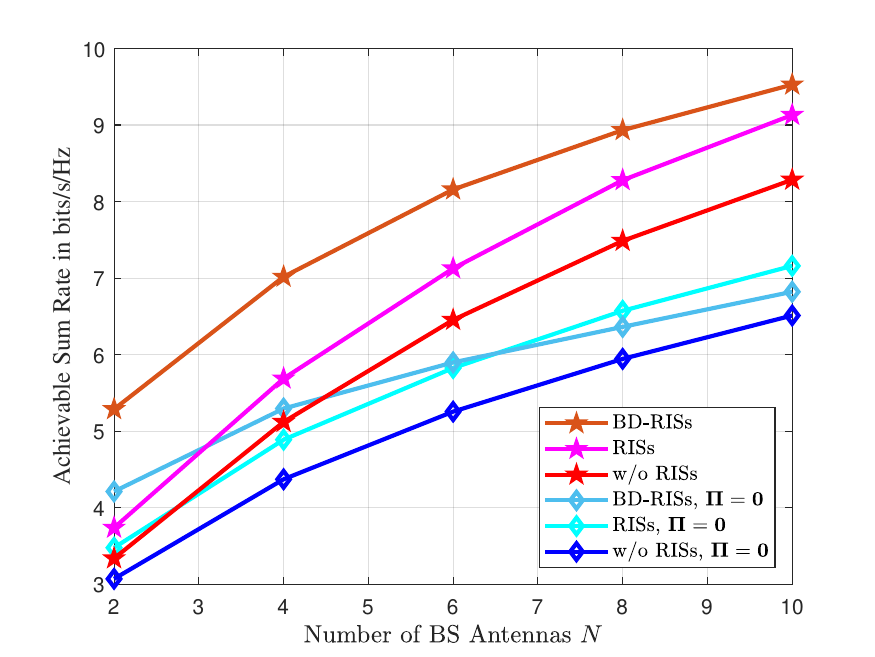}
	\caption{\small{Achievable sum-rate performance as a function of the number $N$ of BS antennas for each user for all investigated schemes, considering $Q=2$ users, transmit power $P=30$ dBm, and $M=100$ elements per RIS.}}
	\label{fig:Rates_vs_N}
\end{figure}

Finally, in Fig.~\ref{fig:Rates_vs_N}, the achievable sum rate is depicted with respect to the common number $N$ of each user's BS antenna array, considering $Q=2$ users, transmit power $P=30$ dBm, and $M=100$ elements per RIS. As observed, the sum rate exhibits a non-decreasing trend for all investigated schemes. Interestingly, for $N \geq 6$, the achievable gains witnessed with the cooperation schemes are much higher than those with the non-cooperative ones, showcasing the benefits of the proposed distributed cooperative design. It can be seen that, as $N$ increases, the gap between the schemes with BD RISs and those with diagonal RISs gets smaller when user cooperation is adopted, while in cases where cooperation is absent and $N>6$, the schemes with diagonal RISs slightly outperform those with BD RISs. However, by taking into account that, as $M$ increases the performance superiority of the schemes with BD RISs increases (as seen in Fig.~\ref{fig:Rates_vs_M}), combined with the fact that an RIS with $M = 100$ elements is relatively small, leads to the intuition that schemes with BD RISs with increasing $N$ and $M$ will outperform their counterparts with diagonal RISs.

\section{Conclusions} \label{Sec:Conclusion}
In this paper, we considered a multiple access system with OFDM, comprising multi-antenna BSs, BD RISs, and single-antenna UEs, and presented a novel distributed framework targeting the system's achievable sum-rate maximization. We focused on the case where each RIS admits a frequency-dependent profile and presented a wideband BD structure to further enhance its design flexibility and resulting performance. The proposed joint design objective includes the precoders at the multiple BSs as well as the tunable capacitances and the switch selection matrices at the multiple BD RISs. A low-complexity and provably convergent algorithm for the distributed optimization of the latter parameters, via minimal message exchange among the deployed BSs, was devised. Our extensive performance evaluation results showcased that the proposed distributed cooperative design is capable of achieving higher sum rates compared to non-cooperative schemes, as well as the importance of taking into account the frequency selective behavior of realistic RIS responses. In addition, the design flexibility of BD RISs for the multi-access system at hand over their diagonal counterparts was numerically investigated. In the future, we intend to extend the system model to the multi-RIS-empowered interference broadcast channel including multi-antenna UEs and study efficient distributed designs under imperfect channel state availability.

\appendices
\section{Proof of Proposition \ref{prop:RIS_Freq_Response}} \label{appx:RIS_Freq_Response}
The transfer function $\phi_{mq}(f,C_{mq})$ for the equivalent resonant circuit of each $m$-th unit element of each $q$-th RIS in \eqref{eqn:reflect_coeff} can be re-written as $\phi_{mq}(f,C_{mq}) = 1 - \phi_{mq}'(f,C_{mq})$, where:
\begin{equation}\label{re}
	\phi_{mq}'(f,C_{mq}) \triangleq \frac{2\mathcal{Z}_0}{\mathcal{Z}(f,C_{mq}) + \mathcal{Z}_0}.
\end{equation}
By replacing with the expression for $\mathcal{Z}(f,C_{mq})$ in \eqref{eqn:characteristic_impedance} and then performing straightforward algebraic manipulations to express the transfer function into a standard form \cite{basso2016linear}, expression \eqref{re} can be simplified as follows:
\begin{equation}
	\phi_{mq}'(f,C_{mq}) = \frac{2}{1 + \frac{\mathcal{D}_{mq}(f,C_{mq})}{\mathcal{N}_{mq}(f,C_{mq})}},
\end{equation}
which concludes the proof.

\section{Proof of Lemma \ref{lem:logarithmic_surrogate}} \label{appx:logarithmic_surrogate}
We first note that $\breve{\mathcal{R}}_{qk}(\vect{w}_{qk})$ can be equivalently re-written, after some algebraic manipulations, as follows:
\begin{align}
	\breve{\mathcal{R}}_{qk}(\vect{w}_{qk}) &= \log_2\left( 1 + \frac{\lvert \vect{f}_{qq,k}^{\rm H}\vect{w}_{qk} \rvert^2}{\operatorname{MUI}_{qk}^t} \right) \\
	&= -\log_2\left( 1 - \frac{\lvert \vect{f}_{qq,k}^{\rm H}\vect{w}_{qk} \rvert^2}{\lvert \vect{f}_{qq,k}^{\rm H}\vect{w}_{qk} \rvert^2 + \operatorname{MUI}_{qk}^t} \right) \\
	&= -\log_2\left(1 - c_{qk}^{-1}|d_{qk}|^2\right), \label{eqn:log_equiv_expr}
\end{align}
where $c_{qk} \triangleq \lvert d_{qk} \rvert^2 + \operatorname{MUI}_{qk}^t$ and $d_{qk} \triangleq \vect{f}_{qq,k}^{\rm H}\vect{w}_{qk}$. Then, it holds that the Right-Hand-Side (RHS) in \eqref{eqn:log_equiv_expr} is a jointly convex function with respect to $(c_{qk},d_{qk})$ \cite{tam2016_successive_all}, which can be lower-bounded by its first-order Taylor expansion $\widehat{\mathcal{R}}_{qk}(c_{qk},d_{qk})$ around the feasible points $(c_{qk}^t$, $d_{qk}^t)$, yielding:
\begin{equation}
	\begin{aligned}
		\breve{\mathcal{R}}_{qk}(c_{qk},d_{qk}) \geq \widehat{\mathcal{R}}_{qk}(c_{qk},d_{qk};c_{qk}^t,d_{qk}^t).
	\end{aligned}	
\end{equation}
Specifically, the following expansion is deduced:
\begin{equation} \label{eqn:taylor_expansion}
	\begin{aligned}
		\widehat{\mathcal{R}}_{qk} =& \breve{\mathcal{R}}_{qk}(c_{qk}^t,d_{qk}^t) + \frac{\partial \breve{\mathcal{R}}_{qk}}{\partial c_{qk}}\Bigg\vert_{\substack{c_{qk} = c_{qk}^t,\\d_{qk} = d_{qk}^t}}(c_{qk} - c_{qk}^t) \\
		&+ \frac{\partial \breve{\mathcal{R}}_{qk}}{\partial d_{qk}}\Bigg\vert_{\substack{c_{qk} = c_{qk}^t,\\d_{qk} = d_{qk}^t}}(d_{qk} - d_{qk}^t)\\ 
		&+ \frac{\partial \breve{\mathcal{R}}_{qk}}{\partial d_{qk}^*}\Bigg\vert_{\substack{c_{qk} = c_{qk}^t,\\d_{qk}^* = (d_{qk}^*)^t}}(d_{qk}^* - (d_{qk}^*)^t),
	\end{aligned}
\end{equation}
where it can be shown, after straightforward manipulations, that the partial derivatives in the RHS of \eqref{eqn:taylor_expansion} are given by
\begin{align}
	\frac{\partial \breve{\mathcal{R}}_{qk}}{\partial c_{qk}} &= -\frac{1}{\ln(2)}\frac{\lvert d_{qk} \rvert^2}{c_{qk}(c_{qk} - \lvert d_{qk} \rvert^2)}, \label{eqn:partial_1} \\
	\frac{\partial \breve{\mathcal{R}}_{qk}}{\partial d_{qk}} &= \frac{1}{\ln(2)}\frac{1}{1 - c_{qk}^{-1}\lvert d_{qk} \rvert^2} c_{qk}^{-1} d_{qk}^*, \label{eqn:partial_2} \\
	\frac{\partial \breve{\mathcal{R}}_{qk}}{\partial d_{qk}^*} &= \frac{1}{\ln(2)}\frac{1}{1 - c_{qk}^{-1}\lvert d_{qk} \rvert^2} c_{qk}^{-1} d_{qk}. \label{eqn:partial_3}
\end{align}
Then, we substitute \eqref{eqn:partial_1}--\eqref{eqn:partial_3} into the RHS of \eqref{eqn:taylor_expansion} and replace properly $c_{qk}$, $d_{qk}$, $c_{qk}^t$ and $d_{qk}^t$, according to their definitions. Finally, using the expressions for $a_{qk}^t$ and $\vect{b}_{qk}^t$ defined in \eqref{eqn:surrog_a} and \eqref{eqn:surrog_b}, omitting the resulting constants, and performing straightforward algebraic manipulations, it can be shown that $\widehat{\mathcal{R}}_{qk}$ is derived as presented in \eqref{eqn:logarithmic_surrogate}, which concludes the proof.

\section{Proof of Theorem \ref{thm:RIS_vectors}} \label{appx:RIS_vectors}
To acquire the expression for $\vect{\gamma}_{\vect{c}_{q}}^t$ ($\underline{\vect{\pi}}_{\vect{c}_q}^t$), it suffices to compute the partial derivative of $\mathcal{R}_q$ ($\mathcal{R}_j$) with respect to ${\vect{c}_{q}}$, according to the definitions in \eqref{eqn:first_surrogate} (in \eqref{eqn:pricing_def}), applying the chain rule. To this end, we first target on the computation of \eqref{eqn:gamma_RIS}. Thus, it can be readily verified that:
\begin{equation} \label{eqn:partial_wrt_phi}
	\frac{\partial \mathcal{R}_q}{\partial \vect{\phi}_{q,k}} = \frac{1}{\ln(2)} \sum_{k=1}^K \frac{1}{(1 + \operatorname{snr}_{qk}^t)\operatorname{MUI}_{qk}^t} \frac{\partial}{\partial \vect{\phi}_{q,k}}\left( \left|\vect{f}_{qq,k}^{\rm H} \vect{w}_{qk}\right|^2 \right).
\end{equation}
By unfolding the term $\left|\vect{f}_{qq,k}^{\rm H} \vect{w}_{qk}\right|^2$, it can be shown that:
\begin{equation*} \label{eqn:ch_f_unfolded}
	\left|\vect{f}_{qq,k}^{\rm H} \vect{w}_{qk}\right|^2 \!=\! \underbrace{\trace\left(\vect{\Phi}_{q,k}^{\rm H}\vect{B}_{qq,k}\vect{\Phi}_{q,k}\vect{C}_{qq,k}\right)}_{\triangleq f_1}\!+\!\underbrace{2\Re\left\{\trace\left(\vect{A}_{qq,k} \vect{\Phi}_{q,k}\right)\right\}}_{\triangleq f_2},
\end{equation*}
where the terms that do not involve the desired variables (i.e.,  $\vect{c}_{q}$) are omitted for brevity. Based on the above expression, we first note that $f_1$ is a quadratic term with respect to $\vect{\Phi}_{q,k}$, while $f_2$ is a linear one. To proceed, we will emphasize on $f_1$ since $f_2$ can be treated in a similar way. Applying the chain rule for $f_1$, yields the following expression for its differential:
\begin{equation}
	\begin{aligned} \label{eqn:differential_f_1}
		{\rm d}f_1 =& \frac{\partial f_1}{\partial (\operatorname{vec}(\vect{\Phi}_{q,k}^*))^{\rm T}}\frac{\partial (\operatorname{vec}(\vect{\Phi}_{q,k}^*))}{\partial\vect{c}_q^{\rm T}}{\rm d}\vect{c}_q \\
		&+ \frac{\partial f_1}{\partial (\operatorname{vec}(\vect{\Phi}_{q,k}))^{\rm T}}\frac{\partial (\operatorname{vec}(\vect{\Phi}_{q,k}))}{\partial\vect{c}_q^{\rm T}}{\rm d}\vect{c}_q.
	\end{aligned}    
\end{equation}
Then, considering the identity $\trace(\vect{A}\vect{B}) = \operatorname{vec}^{\rm T}(\vect{A})\operatorname{vec}(\vect{B})$ as well as trace's cyclic property, we first obtain that:
\begin{align}
	\frac{\partial f_1}{\partial (\operatorname{vec}(\vect{\Phi}_{q,k}^*))^{\rm T}} &= \operatorname{vec}^{\rm T}(\vect{B}_{qq,k}\vect{\Phi}_{q,k}\vect{C}_{qq,k}), \label{eqn:differential_part1_1} \\
	\frac{\partial f_1}{\partial (\operatorname{vec}(\vect{\Phi}_{q,k}))^{\rm T}} &= \operatorname{vec}^{\rm T}(\vect{B}_{qq,k}^{\rm T}\vect{\Phi}_{q,k}^*\vect{C}_{qq,k}^{\rm T}). \label{eqn:differential_part2_1}
\end{align}
However, since $\vect{\Phi}_{q,k}$ is a diagonal matrix, we have that $\operatorname{vec}({\rm d}\vect{\Phi}_{q,k}) = \vect{L}_{\rm d}\operatorname{vec}({\rm d}\vect{\phi}_{q,k})$, where $\vect{L}_{\rm d} \in \mathbb{Z}_2^{M^2 \times M}$ (with $\mathbb{Z}_n\triangleq\{0,1,\ldots,n-1\}$) is a matrix whose role is to place the diagonal entries of an arbitrary square matrix $\vect{A}\in \mathbb{C}^{n\times n}$ on the vector $\operatorname{vec}(\vect{A})$, that is, it holds that: $\operatorname{vec}(\vect{A}) = \vect{L}_{\rm d}\operatorname{vec}_{\rm d}(\vect{A})$ \cite[Definition 2.12]{hjorungnes2011complex}. Also, for the rest of the terms in \eqref{eqn:differential_f_1}, it suffices to again note that $\vect{\Phi}_{q,k}$ is a diagonal matrix and that each $m$-th RIS element's response is independent from the others, i.e., $m' \in \mathcal{M}\triangleq\{1,2,\ldots,M\} \setminus m$. Therefore, it holds that $\frac{\partial (\operatorname{vec}(\vect{\Phi}_{q,k}^*))}{\partial\vect{c}_q^{\rm T}} = \diag\left\{\frac{\partial([\vect{\phi}_{q,k}]_1)^*}{\partial C_{1q}},\ldots,\frac{\partial([\vect{\phi}_{q,k}]_M)^*}{\partial C_{Mq}}\right\}$. Substituting, all above together into \eqref{eqn:differential_f_1}, using the identities $\vect{L}_{\rm d}^{\rm T} \operatorname{vec}(\vect{A}) = \operatorname{vec}_{\rm d}(\vect{A})$ \cite[Lemma 2.24]{hjorungnes2011complex} and $\vect{L}_{\rm d}^{\rm T}\vect{L}_{\rm d} = \vect{I}_M$, as well as the property $\Re\{\vect{Z}\} = \frac{1}{2}(\vect{Z} + \vect{Z}^{\rm H})$ for a Hermitian matrix $\vect{Z}$, the following expression is deduced: 
\begin{equation} \label{eqn:grad_f_1}
	\begin{aligned}
		\nabla_{\vect{c}_q} f_1 = 2\Re&\Bigg\{\diag\left\{\frac{\partial([\vect{\phi}_{q,k}]_1)^*}{\partial C_{1q}},\ldots,\frac{\partial([\vect{\phi}_{q,k}]_M)^*}{\partial C_{Mq}}\right\} \\
		&\hspace{0.3cm}\times \operatorname{vec}_{\rm d}\left( \vect{C}_{qq,k}^{\rm T}\vect{\Phi}_{q,k}^t\vect{B}_{qq,k}^{\rm T} \right)\Bigg\},
	\end{aligned}
\end{equation}
where \cite[eq. (3.2.22)]{Zhang_2017} was used. Following similar steps for the term $f_2$, it can be verified that
\begin{equation} \label{eqn:grad_f_2}
    \begin{aligned}
        \nabla_{\vect{c}_q} f_2 = 2\Re&\Bigg\{\diag\left\{\frac{\partial([\vect{\phi}_{q,k}]_1)^*}{\partial C_{1q}},\ldots,\frac{\partial([\vect{\phi}_{q,k}]_M)^*}{\partial C_{Mq}}\right\} \\ &\hspace{0.3cm}\times \operatorname{vec}_{\rm d}(\vect{A}_{qq,k}^{\rm H})\!\Bigg\}\!.
    \end{aligned}
\end{equation}
Replacing \eqref{eqn:grad_f_1} and \eqref{eqn:grad_f_2} in \eqref{eqn:partial_wrt_phi}, \eqref{eqn:gamma_RIS} can be trivially derived. The same steps apply in the case of the pricing vector in \eqref{eqn:pricing_RIS}, which concludes the proof.

\section{Proof of Corollary \ref{thm:Sel_Mat_vectors}} \label{appx:Sel_Mat_vectors}
Focusing on the derivation of matrix $\vect{\Gamma}_{\vect{S}_q}$ and similarly to the proof of Theorem~\ref{thm:RIS_vectors}, it suffices to unfold the term $\lvert \vect{f}_{qq,k}^{\rm H} \vect{w}_{qk} \rvert^2$ which yields the expression:
\begin{align*}
	\lvert \vect{f}_{qq,k}^{\rm H} \vect{w}_{qk} \rvert^2 &= \left\lvert \left( \vect{h}_{qq,k} + \vect{g}_{qq,k}^{\rm H}\vect{S}_q\vect{\Phi}_{q,k}\vect{H}_{qq,k} \right)\vect{w}_{qk} \right\rvert^2 \\
	&= \vect{w}_{qk}^{\rm H}\Bigg( \vect{h}_{qq,k}(\vect{h}_{qq,k})^{\rm H} + \vect{h}_{qq,k}\vect{g}_{qq,k}^{\rm H}\vect{S}_q\vect{\Phi}_{q,k}\vect{H}_{qq,k} \\ &+\vect{H}_{qq,k}^{\rm H}\vect{\Phi}_{q,k}^{\rm H}\vect{S}_q^{\rm T}\vect{g}_{qq,k}(\vect{h}_{qq,k})^{\rm H} \\ &+\vect{H}_{qq,k}^{\rm H}\vect{\Phi}_{q,k}^{\rm H}\vect{S}_q^{\rm T}\vect{g}_{qq,k}\vect{g}_{qq,k}^{\rm H}\vect{S}_q\vect{\Phi}_{q,k}\vect{H}_{qq,k} \Bigg)\vect{w}_{qk} \\
	&\stackrel{(a)}{=} \underbrace{\trace(\vect{S}_q^{\rm T}\vect{G}_{qq,k}\vect{S}_q\vect{K}_{qq,k})}_{\triangleq g_1} + \underbrace{2\Re\{\trace(\vect{S}_q^{\rm T}\vect{F}_{qq,k}^{\rm H})\}}_{\triangleq g_2},
\end{align*}
where $(a)$ follows from the matrix definitions in \eqref{eqn:S_q_matrices_1}-\eqref{eqn:S_q_matrices_4}, the usage of trace's cyclic property and the neglection of the terms that do not involve $\vect{S}_q$. Then, it can be shown that the partial derivatives of $g_1$ and $g_2$ in the latter expression with respect to $\vect{S}_q^{\rm T}$ are equal to \cite{Zhang_2017}:
\begin{align*}
	\frac{\partial g_1}{\partial \vect{S}_q^{\rm T}} &= \vect{K}_{qq,k}\vect{S}_q^{\rm T}\vect{G}_{qq,k} + \vect{K}_{qq,k}^{\rm H}\vect{S}_q^{\rm T}\vect{G}_{qq,k}^{\rm H},\\
	\frac{\partial g_2}{\partial \vect{S}_q^{\rm T}} &= \vect{F}_{qq,k}^*,
\end{align*}
where the linearity of the trace and the real part operators has been used. However, $\vect{K}_{qq,k} = \vect{K}_{qq,k}^{\rm H}$ and $\vect{G}_{qq,k} = \vect{G}_{qq,k}^{\rm H}$ according to their definitions in \eqref{eqn:S_q_matrices_3} and \eqref{eqn:S_q_matrices_4}, hence, $\frac{\partial g_1}{\partial \vect{S}_q^{\rm T}} = 2\vect{K}_{qq,k}\vect{S}_q^{\rm T}\vect{G}_{qq,k}$. Then, the expression for the gradient (computed at the point $\vect{S}_q^t$) in \eqref{eqn:gamma_Sel_Mat} can be obtained. Performing similar steps, the expression for the pricing matrix $\vect{\Pi}_{\vect{S}_q}^t$ can be derived as well, and thus, the proof is complete.

\section{Proof of Theorem \ref{thm:KKT_Proof}} \label{appx:KKT_Proof}
The proof of Theorem~\ref{thm:KKT_Proof} follows from arguments in the proof of \cite[Theorem 3]{scutari2013decomposition_all}. To this end, we need to assure that the following four assumptions of \cite[Section II]{scutari2013decomposition_all} hold intact:
\begin{enumerate}
    \item The feasible set $\mathcal{X}_q$ is closed and convex.
    \item The function $f_q(\vect{X}_q) \triangleq -\mathcal{R}_q(\vect{X}_q)$ is continuously differentiable on $\mathcal{X}_q$ $\forall$$q$.
    \item Each $\nabla_{\vect{X}_q} f_q(\vect{X}_q)$ has Lipschitz conjugate-gradient on $\mathcal{X}_q$ with constant $\hat{L}$.
    \item The overall sum objective function, that is, $f \triangleq \sum_{q=1}^Q f_q$ is coercive with respect to the feasible set $\mathcal{X} \triangleq \bigcap_{q=1}^Q \mathcal{X}_q$.
\end{enumerate}
Assumptions $2$) and $3$) hold trivially. Regarding assumption $1$), it can be observed that the feasible set $\mathcal{X}_q$ can be constructed as: $\mathcal{X}_q \triangleq \mathcal{W}_q\cap\mathcal{C}_q\cap\mathcal{S}_q$, where $\mathcal{W}_q$ and $\mathcal{C}_q$ represent the feasible sets for the BS linear precoding vectors $\vect{w}_{q}$ and the BD RIS capacitances vectors $\vect{c}_q$, respectively. Also, $\mathcal{S}_q$ is the feasible set for the selection matrix $\vect{S}_q$ at each $q$-th BD RIS. Clearly, $\mathcal{W}_q$ and $\mathcal{C}_q$ are convex sets, while the same holds for $\mathcal{S}_q$ since, by definition, every permutation matrix is a doubly stochastic matrix. Then, Birkhoff's theorem (see, e.g., \cite[Theorem 2.18]{burkard2012assignment}), which states that the Birkhoff polytope (also called the polytope of doubly stochastic matrices) is a convex polytope, yields that $\mathcal{X}_q$ satisfies assumption 1); this happens because the intersection of convex sets remains convex \cite{Boyd_2004}. Finally, assumption $4$), which is equivalent to the statement that all level sets of $f$ are compact, is valid since $\mathcal{X}_q$ is bounded. Overall, the validity of the above assumptions indicates that \cite[Lemma 6]{scutari2013decomposition_all} holds true, which further implies that the following two conditions are satisfied:
\begin{align}
    &(\vect{X}\!-\!\vect{U})^{\rm H}\left( \nabla_{\vect{X}}f(\vect{X};\vect{Y})\!-\! \nabla_{\vect{U}}f(\vect{U};\vect{Y}) \right) \geq c_{\tau} \norm{\vect{X}\!-\!\vect{U}}_{\rm F}^2, \label{eqn:strongly_convex} \\
    &\norm{\nabla_{\vect{X}}f(\vect{X};\vect{Y}) - \nabla_{\vect{U}}f(\vect{X};\vect{U})}_{\rm F} \leq B\norm{\vect{X} - \vect{U}}_{\rm F}, \label{eqn:Lipschitz_continuous}
\end{align}
with $0<B<\infty$ and for all $\vect{X}, \vect{U} \in \mathcal{X}_q$, as well as for a given $\vect{Y}\in \mathcal{X}_q$. In other words, \eqref{eqn:strongly_convex} and \eqref{eqn:Lipschitz_continuous} mean that $f$ is uniformly strongly convex with constant $c_{\tau}$, and that $\nabla_{\vect{X}}f(\vect{X})$ is uniformly Lipschitz continuous on $\mathcal{X}_q$, respectively.

It can be deduced from the latter assumptions and conditions that \cite[Proposition 1]{scutari2013decomposition_all} holds, which plays a pivotal role for the proof of Theorem~\ref{thm:KKT_Proof}. In fact, based on standard descent arguments suitably combined with the properties of the best-response mapping $\widehat{\vect{X}}_q^t$, the proof can be readily concluded. At this point, it is also important to note that each design variable included in $\vect{X}_q$, i.e., $\vect{w}_q^t, \vect{c}_q^t$ and $\vect{S}_q^t$, is derived as a global optimum by solving the corresponding sub-problems $\mathcal{OP}_2$, $\mathcal{OP}_3$, and $\mathcal{OP}_4$. In particular, to obtain $\vect{w}_q^t$, we solve the relaxed problem $\widehat{\mathcal{OP}}_2$ in closed form, while preserving the first-order properties (cf. \eqref{eqn:obj_surrog_gradients}) of $\mathcal{OP}_2$'s objective function, due to the Taylor expansion in Lemma \ref{lem:logarithmic_surrogate}. Moreover, $\mathcal{OP}_3$ is solved in closed form and $\mathcal{OP}_4$'s global solution can be derived via the total unimodularity property. To proceed, we first note that, due to the Descent Lemma \cite{bertsekas1999nonlinear}, it holds for any given $t\geq 0$ that:
\begin{equation} \label{eqn:condition1}
\begin{aligned}
    f(\vect{X}^{t+1}) \leq f(\vect{X}^t) &+ \alpha^t\nabla_{\vect{X}}f(\vect{X}^t)^{\rm H}(\widehat{\vect{X}}^t - \vect{X}^t)\\
    &+ \frac{(\alpha^t)^2 \hat{L}}{2} \left\|\widehat{\vect{X}}^t - \vect{X}^t \right\|_{\rm F}^2.
\end{aligned}
\end{equation}
In addition, by using \cite[Proposition 1.(c)]{scutari2013decomposition_all}, which in our case reads as follows:
\begin{equation} \label{eqn:condition2}
    \nabla_{\vect{X}}f(\vect{X}^t)^{\rm H}(\widehat{\vect{X}}^t - \vect{X}^t) \leq -\tau\left\| \widehat{\vect{X}}^t - \vect{X}^t \right\|_{\rm F}^2,
\end{equation}
we obtain, after combining \eqref{eqn:condition1} and \eqref{eqn:condition2} and using the property $\alpha^t \rightarrow 0$ for sufficiently large $t$, the following inequality:
\begin{equation} \label{eqn:condition3}
    f(\vect{X}^{t+1}) \leq f(\vect{X}^t) - \alpha^t\left(\tau - \frac{(\alpha^t) \hat{L}}{2}\right) \left\|\widehat{\vect{X}}^t - \vect{X}^t \right\|_{\rm F}^2.
\end{equation}
This inequality indicates that $f(\vect{X}^{t+1})$ is a decreasing sequence of values, implying that no limit point of $\{\vect{X}^t\}_{t=0}^\infty$ can be a local maximum for $f$; this completes the proof.

\bibliographystyle{IEEEtran}
\bibliography{refs}

\end{document}